\documentclass[useAMS,usenatbib]{mn2e}
\usepackage{graphicx}
\begin{document}

\title[Stellar Populations in Bulges of Spiral Galaxies]{Stellar Populations in 
Bulges of Spiral Galaxies\thanks{Based on observations obtained with the Apache 
Point Observatory 3.5-meter telescope, which is owned and operated by the 
Astrophysical Research Consortium.}}

\author[Bhasker K. Moorthy \& Jon A. Holtzman]{Bhasker K. Moorthy\thanks{E-mail:
bmoorthy@nmsu.edu} \& Jon A. Holtzman\\Department of Astronomy, Box 30001, MSC 4500, 
New Mexico State University,Las Cruces, NM 88003}

\date{Accepted 0000 month 00. Received 0000 month 00; in original form 
0000 month 00}

\pagerange{\pageref{firstpage}--\pageref{lastpage}} \pubyear{0000}

\maketitle

\label{firstpage}

\begin{abstract}
We present line strengths in the bulges and inner disks of 38 galaxies in the 
local universe, including several galaxies whose bulges were previously identified 
as being disk-like in their colors or kinematics, to see if their spectral properties 
reveal evidence for secular evolution.  We find that red bulges of all Hubble types 
are similar to luminous ellipticals in their central stellar populations.  They have 
large luminosity-weighted ages, metallicities, and $\alpha$/Fe ratios.  Blue bulges 
can be separated into a metal-poor class that is restricted to late-types with small 
velocity dispersion and a young, metal-rich class that includes all Hubble types and 
velocity dispersions.  Luminosity-weighted metallicities and $\alpha/Fe$ ratios are 
sensitive to central velocity dispersion and maximum disk rotational velocity.  Red 
bulges and ellipticals follow the same scaling relations.  We see differences in some 
scaling relations between blue and red bulges and between bulges of barred and 
unbarred galaxies.  Most bulges have decreasing metallicity with increasing radius; 
galaxies with larger central metallicities have steeper gradients.  Where positive 
age gradients (with the central regions being younger) are present, they are 
invariably in barred galaxies.  The metallicities of bulges are correlated with those 
of their disks.  While this and the differences between barred and unbarred galaxies 
suggest that secular evolution cannot be ignored, our results are generally 
consistent with the hypothesis that mergers have been the dominant mechanism 
responsible for bulge formation.  \end{abstract}

\section{Introduction}

%scat light corr for red channel

Bulges are important relics of the galaxy formation process.  An analysis of their 
structure, kinematics, dynamics, and stellar content can potentially reveal the 
physical mechanisms responsible for the formation and evolution of galaxies as 
well as the nature of the Hubble sequence.  Similarities between bulges and 
ellipticals have long been recognized but recent observations suggest that at 
least some bulges may be related to disks.  This has led to the suggestion that the 
large bulges of early-type spirals are more similar to ellipticals while late-type 
bulges are more disk-like \citep{1997ARA&A..35..637W}.  As a consequence of these 
observations, formation scenarios have emerged for bulges that are either identical 
to those for ellipticals or involve the secular evolution of disks.  However, the 
degree to which formation mechanisms are homogeneous is still open to question.

Early models for elliptical formation involved the monolithic collapse of a 
primordial gas cloud \citep{1974MNRAS.166..585L, 1984ApJ...286..403C, 
1987A&A...173...23A}.  This model naturally explains several observed properties of 
ellipticals including the mass-metallicity relation and the presence of metallicity 
gradients but large-scale collapse is inconsistent with present day cold dark matter 
cosmology and with recent observations showing that massive ellipticals were not 
fully assembled since after z$=$1 \citep{2004ApJ...608..752B, 2005astro.ph..6044F}.  
It is now widely believed that ellipticals formed hierarchically through mergers of 
smaller fragments \citep{1993MNRAS.264..201K}.  Mergers are frequently caught in the 
act \citep{1999ApJ...520L..95V, 2004A&A...428..837F}  and photometric and kinematic 
evidence for past mergers is abundant in ellipticals \citep{2004MNRAS.352..721E, 
2005astro.ph..6661V}.  The merger model has been extended to bulges due to the many 
observed similarities between bulges and ellipticals.  For 
example,\cite{1997AJ....114.2366C} found that bulges were well-fit by the R$^{1/4}$ 
law used for ellipticals.  The fundamental plane relation of bulges is nearly the 
same as that of ellipticals, with late-types perhaps lying below early types 
\citep{2002MNRAS.335..741F}.  

% sec evol theory and obs
In the secular evolution scenario, bulges are produced through radial and vertical 
transport of disk material as the result of instabilities and resonances (see 
Kormendy \& Kennicutt 2004 for a review).  These models come in several flavors, most 
of which involve bars.  Simulations that do not include gas have found that bars can 
buckle, heating the inner disk and increasing its scale height to resemble a bulge.  
Hydrodynamical simulations have found that bars can transport gas towards the center, 
triggering intense star-formation \citep{1993gabu.symp..387P, 1995A&A...301..649F, 
1996ApJ...462..114N, 2000MNRAS.312..194N, 2004A&A...413..547I}.  The presence of 
neighbors may also drive this \citep{2004AJ....127.1371K}.  Support for secular 
evolution comes from observed correlations between the scale lengths of bulges and 
their disks \citep{1996ApJ...457L..73C, 2003ApJ...582..689M}.  Recent work has also 
shown that the light profiles of many bulges are closer to exponential than R$^{1/4}$ 
\citep{2003ApJ...582L..79B, 2003ApJ...582..689M,2004MNRAS.355.1155D}.  Furthermore, 
the ratio of rotational to random motions in bulges is often typical of disks 
\citep{1982ApJ...256..460K, 2004ARA&A..42..603K}.  Comparisons between the morphology 
and kinematics of observed galaxies with simulated ones have shown that boxy and 
peanut-shaped (b/p) bulges are bars viewed at high inclination 
\citep{1999AJ....118..126B, 2003Ap&SS.284..753A, 2004AJ....127.3192C, 
2005MNRAS.358.1477A}.  Athanassoula (2005) distinguishes between b/p bulges, which 
are formed through the buckling of the bar, and what she calls ``disky bulges", which 
are smaller cold components that formed out of the gas driven inward by the bar.

Bulges that could have been formed through secular evolution are often referred to as 
``pseudobulges" to distinguish them from the ``classical" bulges that may have formed 
through mergers.  Since pseudobulge signatures are generally found in later-typed 
spirals, Kormendy \& Kennicutt (2004) suggest that early-type spirals (Sa's, Sab's, 
and some S0's) contain classical bulges while late-type spirals (Sb's, Sc's and some 
S0's) contain pseudobulges.  On the theoretical side, \cite{1993gabu.symp..387P} 
found that secular evolution can produce small bulges but not those having a 
characteristic radius much larger than the disk scale length.  However, it is not at 
all clear that the spectrum of observed bulge properties points towards two distinct 
formation scenarios.  Since the stability of bars continues to be debated 
\citep{2004ApJ...604..614S, 2004ApJ...604L..93D, 2005MNRAS.tmpL..89B}, it is also not 
clear whether or not pseudobulges should exist only in present-day barred galaxies.

Stellar population (SP) studies can potentially place important constraints on the 
formation mechanisms.  A succesful formation scenario has to reproduce the observed 
distribution of ages and metallicities.  In a collapse model, bulges and ellipticals 
are universally old and have radial metallicity gradients.  In his dissipative 
collapse simulation \cite{1984ApJ...286..403C} found that the steepness of the 
metallicity gradient was correlated with galaxy properties such as mass and 
luminosity.  If ellipticals and bulges formed through mergers, it is important to 
keep in mind that their assembly histories might be very different from their
star formation histories.  $\Lambda$CDM simulations suggest that most
massive ellipticals (and therefore presumably bulges) were not fully assembled until
recently (z$<$1) whereas the bulk of star formation occurred much earlier (z$>$2) in 
the progenitor galaxies \citep{2005astro.ph..9725D}.  This is consistent with 
observational studies of merger activity, number counts, and the luminosity function 
\citep{2005astro.ph..6044F, 2005astro.ph..6661V}.  de Lucia et al. find that 
the star formation histories of massive ellipticals peak at z$\approx$5 while 
those of less massive ellipticals peak at progressively smaller redshifts and are
more extended.  These simulations predict a mass-metallicity relation, with the most 
massive ellipticals having solar metallicity and the least massive ones being a 
factor or ten smaller in metallicity.  Gradients in SPs are difficult to model 
within the framework of hierarchical formation.  Mergers between disks can presumably 
preserve existing gradients in the subcomponents and produce new gradients through 
gas infall, but mixing from successive mergers might erase any correlations between 
gradients and global properties.  White (1980) found that the metallicity gradient in 
a disk galaxy was halved after three mergers with similar sized disks.  However, 
\cite{1999ApJ...513..108B} found that more massive galaxies had steeper metallicity 
gradients.  The impact of secular evolution on gradients is not straightforward.  
Since the resulting pseudobulge has a smaller scale length than the progenitor disk, 
an existing disk gradient could become amplified (A Klypin, private communication).  
However, mixing during secular evolution can have the effect of washing out existing 
gradients.  Adding gas only complicates the picture.  If gas is fueled towards the 
central regions by bars, this could result in a nucleus that is younger and more 
metal-rich than the outer regions of the bulge.  Simulations by Friedli et al. (1994) 
resulted in a flattening of metallicity gradients in all but the innermost regions 
where a starburst, fueled by infalling gas, produced a metal-rich nucleus.

Abundance ratios can place additional constraints.  Mg and other $\alpha$-elements 
are primarily produced in Type II Supernovae (SN II) while a substantial fraction of 
the Fe-peak elements Fe and Cr are produced in Type Ia Supernovae (SN Ia).  
Therefore, $\alpha$-enhancement is generally attributed to a cessation of star 
formation before the bulk of SN Ia occurred.  Through chemical evolution modeling, 
Thomas et al. (1999) found that a clumpy collapse model produced uniform 
$\alpha$-enhancement or positive gradients (increasing $\alpha$/Fe with radius) while 
a merger model produced uniformly solar $\alpha$/Fe or negative gradients.  For the 
case of secular evolution, \cite{2004A&A...413..547I} make different predictions for 
abundance ratios in bulge stars depending on whether it is the gas disk or the 
stellar disk which first becomes unstable.  In the former case, gas clumps merge 
together and spiral inward, causing massive starbursts and producing large 
$\alpha$/Fe ratios.  In the latter case, a bar forms and then channels gas towards 
the center.  This occurs on long timescales, resulting in smaller $\alpha$/Fe ratios.

The only bulges where individual stars can be resolved are those of the 
Milky Way (MW) and M31.  The majority of the stars in the MW bulge are old (t$\geq$7 
Gyr), although young (t$\leq$200 Myr) and intermediate-age (200 Myr $\leq$t$\leq$7 
Gyr) stars are also detected \citep{1995MNRAS.275..605I, 1996AJ....112..171S, 
2000A&A...355..949F, 2003MNRAS.338..857V, 2003A&A...399..931Z}.  As a barred 
late-type spiral, the MW might be a good candidate for secular evolution but that 
hypothesis is challenged by the mean stellar age of its bulge.  If the bulge were 
produced through a rearrangement of disk stars, this must have occurred several Gyr 
ago if the inner disk has the same age distribution as the disk at the solar 
neighborhood.  The young stars in the bulge are mainly found in the innermost 
regions.  While the gas that formed them could have been driven by a bar, it could 
just as well have been provided by a recent merger.  The mean metallicity of the 
Galactic bulge is slightly sub-solar \citep{1996ApJ...459..175M, 
2000A&A...355..949F,2003MNRAS.338..857V, 2003A&A...399..931Z}(Recent Fullbright et 
al. 2006).  The stellar content of M31's bulge is not as well understood as that of 
the MW since we cannot reach its main sequence turnoff.  Observations of giant stars 
are consistent with M31's bulge being similar in age to the MW bulge and slightly 
more metal-rich\citep{1999ASPC..192..215R, 2000A&A...359..131J, 2001AJ....122.1386D, 
2003AJ....125.2473S, 2005AJ....130.1627S}.

SPs in more distant bulges have to be studied through photometry or spectroscopy of 
integrated light.  An important limitation of such studies is that integrated light 
is dominated by the most luminous stars.  Colors have been studied more extensively 
as they have the advantage of higher signal-to-noise (S/N).  Pioneering work by 
\cite{1994AJ....107..135B} found that color variations from galaxy to galaxy are much 
larger than color differences between disk and bulge in each galaxy.  Similarly, 
\cite{1996A&A...313..377D} found that bulge and disk colors are correlated and that 
the color differences between bulge and disk suggested that the SPs 
did not vary much from one to the other.  Unfortunately, color studies suffer from 
degeneracies between ages, metallicities, and extinction.

Line strengths are nearly insensitive to dust \citep{2005ApJ...623..795M}, provide 
information on the abundances of several elements and molecules, and allow for 
breaking the age-metallicity degeneracy.  \cite{1994ApJS...95..107W} obtained line 
strengths for a range of single age, single metallicity SPs (SSPs) on 
the Lick/IDS system \citep{1984ApJ...287..586B, 1985ApJS...57..711F} and found that 
while individual indices are sensitive to both age and metallicity, the relative 
sensitivity varies from index to index.  Spectral indices have also been defined at 
high resolution \citep{1999ApJ...513..224V}, allowing better age determinations than 
otherwise possible.  One of the limitations of the original models is that they were 
calibrated using galactic stars, few of which had abundance ratios different from 
solar.  Much progress has since been made in extending Lick indices to non-solar 
abundance ratios \citep{1995AJ....110.3035T, 2000AJ....119.1645T, 
2003MNRAS.339..897T, 2004MNRAS.351L..19T, 2005ApJS..160..176L}.

%central line strengths ellipticals
Line strengths have been used extensively to characterize the SPs of 
ellipticals.  The luminosity-weighted ages of cluster ellipticals are large while 
those of field ellipticals are on average smaller, with a large spread 
\citep{1994AJ....108.2054R, 2000AJ....119.1645T, 
2001ApJ...551L.127V, 2002MNRAS.333..517P, 2005MNRAS.358..813D, 2005ApJ...621..673T}.  
This goes against the collapse model and confirms, at least qualitatively, the 
prediction of the merger model by Kauffmann (1996).  Mg-sensitive indices in 
ellipticals are more tightly correlated with central velocity dispersion than 
Fe-sensitive indices, resulting in a correlation between Mg/Fe and $\sigma_0$ 
\citep{1993ApJ...411..153B}.  \cite{2003ApJ...586...17W} found that that the 
Mg-$\sigma$ relation of ellipticals is consistent with these objects having been 
formed through around 50 mergers with merger probability constant or midly declining 
with time.

%central line strengths bulges
There have been fewer studies of line strengths in bulges.  Integrated light 
studies on the bulges of the MW and M31 have arrived at similar ages and 
metallicities as the resolved studies \citep{2002A&A...395...45P, 
2005A&A...434..909P}.  Both bulges have large SSP ages.  M31 is slightly super-solar 
in SSP metallicity while the Milky Way is solar.  Both are $\alpha$-enhanced with M31 
being more so in line with its larger $\sigma_0$.  Early studies on extragalactic 
bulges found them to be similar to ellipticals in their central line strengths 
\citep{1996AJ....112.1415J, 1996AJ....112.2541I}.  \cite{2002MNRAS.333..517P} found 
that bulges have smaller average luminosity-weighted age than ellipticals.  These 
authors did not find the correlation predicted by Kauffman (1996) for bulges and 
suggested that it might have been erased by secular evolution in late-types. 
The largest sample of bulges to date was that of \cite{2001A&A...366...68P}, who 
identified three classes of bulges: a) young bulges which are small, have 
ionized gas, low velocity dispersions, and low metallicity; b) old bulges that are 
alpha-enhanced and follow the mass-metallicity relation of ellipticals; and c) bulges 
that have a mixture of young and old populations, which are less alpha-enhanced than 
those of class (b), and deviate from the Mg$_2$ relation of ellipticals.  Prugniel et 
al. and Proctor et al. found that both Fe and Mg were correlated with $\sigma_0$ in 
bulges, resulting in the lack of a tight correlation between Mg/Fe and $\sigma_0$ in 
bulges.  \cite{2001A&A...366...68P} found that Mg$_2$ in bulges is more tightly 
correlated with the V$_{max}$ of the disk than with $\sigma$, indicating that the 
SPs are more sensitive to the total galaxy potential (i.e. the dark 
matter halo) than the bulge potential.

%motivation for gradients
Studies with spatial resolution offer several advantages to studies that only sample 
the central region.  First, differential studies of ages and abundances are more 
reliable than absolute estimates.  Second, formation models invariably make 
predictions for the global properties of galaxies which are better traced by mean 
observed quantities than central ones; observations with spatial resolution allow 
estimation of mean values.  Finally, as mentioned already, population gradients can 
place additional constraints on formation mechanisms.

%grads ellips and bulges
Line strength gradients have been studied extensively in ellipticals.  
\cite{1993MNRAS.265..553C} and \cite{2005MNRAS.tmpL..47F} find strong correlations 
between gradients and physical properties while others find weak (Mehlert et al. 
2003) or no \citep{1999ApJ...527..573K} correlations.  There have been relatively few 
studies on gradients in bulges.  Fisher et al. (1996) found steeper metallicity 
gradients along the minor axes of nine edge-on S0s than along the major axes, 
suggesting different formation mechanisms for the bulge and the disk.  
\cite{1999Ap&SS.269..109G} found that gradients were correlated with luminosity in 16 
bulges.  \cite{2000MNRAS.311...37P} found that gradients correlated with velocity 
dispersion, albeit with a sample of only four galaxies, while 
\cite{2002Ap&SS.281..367J} found no such correlation.   Integral field spectroscopy 
has enabled the acquisition of 2D line strengths in bulges with results just starting 
to emerge (e.g. Sil'chenko et al. 2003; Falc{\'o}n-Barroso et al. 2004).  Recent work 
by \cite{2005MNRAS.358.1337R} shows that tunable filters might be another way to 
obtain 2D line strengths.

In this paper, we present line strengths and line strength gradients in the bulges 
and inner disks of 38 galaxies.  Our sample, described in Section 2, was chosen to 
span a range of bulge properties and specifically targeted several galaxies with blue 
bulges and similar bulge/disk colors and/or disk-like kinematics in an attempt to 
look for SP signatures of secular evolution.  Section 3 describes the observations 
and data analysis.  Section 4 describes the SP results and Section 5 discusses their 
implications for bulge formation scenarios.  Section 6 contains a summary.  The 
structure, kinematics, and dynamics and how they relate to the SPs will be discussed 
in a future paper (hereafter Paper II).

\section{The Galaxy Sample}

We selected a sample that included some bulges that are similar in color to their 
disks and others that are considerably redder as a control.  Color was chosen as the 
primary selection criterion because this has so far been the best studied property of 
bulges.  \citet[hereafter DJ]{1996A&A...313..377D} and \citet[hereafter 
PB]{1997NewA....1..349P} obtained color gradients of galaxies from the Uppsala 
General Catalog \citep{1973ugcg.book.....N} with major axes larger than 2'.  We 
selected 17 galaxies PB and 14 from DJ.  The two samples complement each other 
nicely in their sky coverage and sampling of Hubble types.  The DJ galaxies are 
nearly face-on while the PB galaxies are highly inclined.

We also included three galaxies, NGCs 2787, 3384, and 3945, which were previously 
found to possess disk-like structural and kinematic properties 
\citep{1996A&A...314...32B, 2003ApJ...591..185S, 2003ApJ...596..903P, 
2003ApJ...597..929E}.  All three are barred S0 galaxies with inner disks or bars that 
are more luminous than the surrounding bulge.  One of the PB galaxies, NGC 7457, is 
also known to have disk-like kinematics \citep{1993IAUS..153..209K, 
2003ApJ...596..903P}.  \cite{1994A&AS..107..187M} found small bar-like distortions in 
this galaxy and \cite{2004MNRAS.352..721E} found nearly cylindrical rotation which is 
seen in boxy bulges \citep{1977ApJ...211..697B, 1982ApJ...256..460K, 
2004MNRAS.350...35F} and in simulations of edge-on bars \citep{1990A&A...233...82C, 
1993IAUS..153..391S, 2002MNRAS.330...35A}.

This project initially began in collaboration with some members of the ENEAR survey 
\citep{2000cofl.work...62W}.  Therefore, the first five galaxies we observed were 
from their sample: NGCs 4472, 2775, 3544, 3831, and 5793.  NGC 4472, a bright 
elliptical in the center of the Virgo cluster, was included for comparison with 
previous studies.  The other four galaxies were selected to span a wide range in 
inclination and bulge-to-disk ratio.

There are several reasons for including both high- and low-inclination galaxies:

1. Minor axis observations of highly inclined galaxies offer minimum disk 
contamination in the outer regions of the bulge.

2.  In moderately inclined galaxies, there is actually more disk contamination along 
the minor axis than the major axis for the same solid angle.  To estimate the degree 
of disk contamination, we obtained spectra along both major and minor axes for some 
of our inclined galaxies.

3. Major axis observations of inclined galaxies allow for the measurement of rotation 
which can provide additional information about the structure of the galaxy.  

4. Low-inclination galaxies have less disk contamination in the central regions 
and allow for clear identification of bars, rings, and other morphological features.  
Including both high- and low-inclination galaxies allows for a comparison between 
bars and b/p bulges.

5. In highly-inclined galaxies, the bulge and disk can be distinguished based on 
their shapes (spheroidal versus flat).  In face-on galaxies, this is not possible and 
so bulges are generally defined as the excess light on top of the inward 
extrapolation of an exponential disk.  SPs and kinematics offer two 
additional and independent means of distinguishing between bulges and disks in 
face-on galaxies.

Twelve out of our 20 low-inclination galaxies are barred.  Some of our 
highly-inclined galaxies were classified by \cite{2000A&AS..145..405L} into 
peanut-shaped, boxy, nearly boxy, or elliptical bulges.  For our remaining 
highly-inclined galaxies, we determined the shapes using their technique.  This 
yielded 10 b/p bulges and 8 elliptical bulges.  Therefore the fraction of barred 
galaxies in low-inclination galaxies is approximately equal to the fraction of b/p 
bulges in highly-inclined galaxies.  While peanut-shaped bulges are easily 
identified, it is not always easy to distinguish an elliptical bulge from one that is 
slightly boxy.  For instance, L{\" u}tticke et al. classify NGC 5838's bulge as 
elliptical but \cite{1994A&AS..107..187M} describe it as boxy.

Table 1 contains basic data on our galaxies.  The column ``Morph." describes the 
shape of the bulge if the galaxy is highly-inclined (Boxy, Peanut, or Elliptical) and 
whether or not it is barred if it is not highly-inclined.  Identifications marked 
with an asterisk are those of L{\" u}tticke et al. while those without asterisks are 
our identifications.

\begin{figure}
{\includegraphics[width =0.45\textwidth]{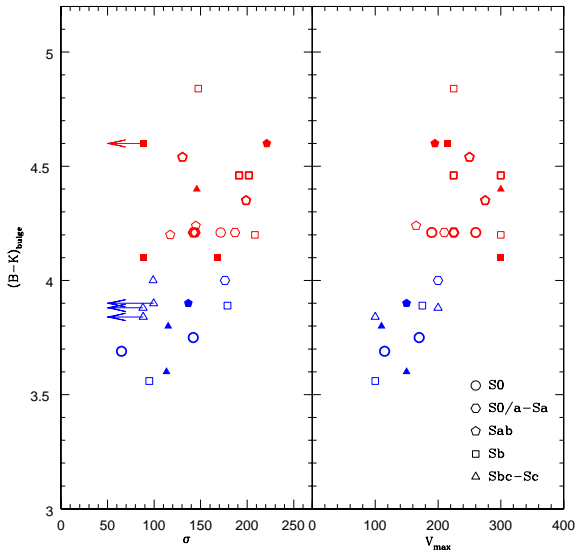}}
 \caption{Bulge color versus central velocity dispersion and maximum disk rotational 
velocity.  The velocities are from this work while the colors are from PB and DJ.  
The central velocity dispersions were measured within an aperture of approximately 4 
arcsec.  See Table 1 for the definition of bulge color.  Bulges are shown in red or 
blue according to whether they are redder or bluer than B-K=4.  Filled symbols 
are barred galaxies.  Thin open symbols are elliptical bulges if highly-inclined and 
unbarred galaxies if not highly-inclined.  Thick open symbols are b/p bulges.}
\label{bksigvmax}
\end{figure}

When comparing SPs in galaxies with different colors, it is important 
to keep in mind that color is correlated with the global dynamical properties of a 
galaxy.  Fig. \ref{bksigvmax} shows the bulge B-K colors as a function of central 
velocity dispersions, and maximum disk rotational velocities of our galaxies where 
available.  We found that it is useful to subdivide bulges according to whether they 
are redder or bluer than B-K=4; these are shown as red and blue points.  In this 
and subsequent plots, point shape represents the Hubble type: circles are S0's; 
hexagons are S0a's and Sa's; pentagons are Sab's; squares are Sb's; and 
triangles are Sbc's and Sc's.  Filled symbols are barred galaxies.  Thin open 
symbols are elliptical bulges if highly-inclined and unbarred galaxies if not 
highly-inclined.  Thick open symbols are b/p bulges.

Bulge colors correlate more tightly with V$_{max}$ than with 
$\sigma_0$.  Galaxies with V$_{max}>200 km s^{-1}$ host red bulges while those with 
V$_{max}<165 km s^{-1}$ host blue bulges.  Both red and blue bulges are found in 
nearly the full range of central velocity dispersions spanned by our galaxies 
although there is an overabundance of red bulges in large $\sigma$ galaxies and vice 
versa.

%[currently using published values for b/a, z, and scale.  state ours instead and if 
%we don't have images, state where you got the b/a]%[replace b-r in table 1 with b-k]

%fix bottom margin in first page of table 1 

\section{Observations and Data Analysis}

\subsection{Observations}

Observations were made with the Double Imaging Spectrograph (DIS) on the ARC 3.5m 
telescope at Apache Point Observatory between January 2000 and February 2004.  The 
spectrograph uses a dichroic to split the light into separate blue and red channels.  
During this period, the instrument was upgraded in several phases with the 
installation of new detectors and optics.  Table 2 gives the specifications of each 
configuration and Table 3 describes the spectroscopic observations.  DIS I gave us 
continuous wavelength coverage from 4000 to 7500 \mbox{\AA} while DIS II and DIS III 
gave us continuous coverage from 3700 to 7500 \mbox{\AA}.  A 5 arcmin x 1.5 arcsec 
slit was used in all the observations.  On each night, we observed a quartz lamp for 
flatfielding, arc lamps for wavelength calibration, and two to five 
spectrophotometric standards for flux calibration. On several nights, Lick standard 
stars from \cite{1994ApJS...94..687W} were observed to allow us to transform our line 
indices to the Lick/IDS system.

% not all pa's incl in table 3

For most of the highly inclined galaxies, we obtained spectra along 
both major and minor axes.  For the unbarred low-inclination galaxies, we obtained 
major axis profiles except for NGC 2916 for which we obtained a minor axis profile 
instead because there was a bright star along the major axis.  For the clearly barred 
galaxies, we placed the slit along the bar.  For IC 302 and NGC 2487, this is 
different from the position angle of the major axis.  In NGC 5375, the bar happens to 
be along the major axis.  We could not identify a major axis in NGC 266 or NGC 5020 
but the slit must have been placed close to it as we see substantial rotation.  We 
did not detect any rotation in five galaxies: IC 302 and NGCs 765, 2487, 2916, and 
6246A, due to their inclinations being too low.

We obtained images in B, V, and R for bulge-to-disk decomposition on six nights using 
the SPIcam detector on the same telescope.  Typical exposure times were 600 sec for 
B, 180 sec for V, and 120 sec for R.  Superbiases and twilight flats were obtained on 
each night.  

%Conditions were photometric on four of the six nights.  The galaxies observed on the 
%non-photometric night are IC 1029 and NGCs 5375, 5707, 5719, 6246A, 6368, 7311, 
%7332, 7457, and 7537.  

%[7311-compare kinem in first night with 2nd night to figure out which closer to maj 
%axis; incl. results for both]

\subsection{Basic Reductions}

Data reduction was carried out using the XVISTA software package.  For the imaging, 
basic reduction included bias subtraction and flatfielding.  For the 
spectroscopy, flat-fields were constructed using a median of 5 
to 10 bright quartz lamp exposures; the mean spectral response was divided out. 
Wavelength calibration was performed using He, Ne, and Ar arc lamp exposures, using a 
fifth order polynomial for both blue and red channels.  Flux calibration was 
performed using a spline fit to published spectra of the spectrophotometric standards 
\citep{1988ApJ...328..315M}.  Line curvature along the slit was measured using the 
lamp exposures and a simple row by row shift was stored for subsequent correction.  
Similarly, spatial distortion in the spectrograph was measured using standard 
star exposures and the correction was stored.  Since spatial distortion includes a 
component due to differential refraction unless the slit is perpendicular to the 
parallactic angle, this component was calculated and removed from the standard star 
measurements.  Application of a correction to the galaxies includes a refraction 
component and the spectrograph component.

Each galaxy frame was subtracted by the overscan and superbias, flat-fielded, and 
trimmed to remove the overscan region.  The line curvature and spatial distortion 
corrections were applied.  Multiple observations of each galaxy were then coadded, 
rejecting cosmic ray outliers in the process.  Where multiple observations were not 
available, cosmic rays were removed by eye using a spatial median filter.  Variance 
frames were propogated throughout the reduction process.  To combine the red and blue 
channels, the red galaxy spectra were rescaled to spatially match the blue spectra.  
The spectra were then extracted in approximately 1 arcsec bins near the center of the 
galaxy and using larger bins further out.  An average of sky values measured on both 
sides of the galaxy was subtracted from each slice. Wavelength and flux calibrations 
were applied and both red and blue frames were rebinned to 3 \AA/pixel. Finally, the 
spectra were deredshifted and the red and blue sides were combined.  The blue 
spectrum was used out to a wavelength of 5600 \mbox{\AA} and the red from 5450 \AA. 
In the overlap region between 5450 and 5600 \AA, an average of red and blue values 
was used. In some galaxies, there is a discontinuity in the overlap region that 
arises from the difficulty of accurately combining the two channels for extended 
objects; fortunately none of our absorption features fall within this region.

On our last observing run (Feb 15, 2004), the light on star frames was too extended 
to be explained by seeing alone.  The excess light, which we believe is 
scattered light, had a spectral energy distribution similar to that of the mashed 
stellar spectrum but without any narrow spectral features presumably because it is 
significantly defocused.  Since adding a constant to a spectrum decreases the 
equivalent width (EW) of an absorption features and since the relative contribution 
of the scattered light to galaxy light increases with distance from the galaxy 
center, this introduces an artificial negative gradient in the line strength 
profiles.  To correct for the scattered light, we fit the 2d stellar spectrum 
with a smoothed stellar spectrum along the wavelength direction and a fifth order 
polynomial along the spatial direction, masking out the central 20 pixels (8.4 
arcsec).  This spatial profile, combined with the smoothed spectral profile 
of the galaxy, was subtracted from each galaxy frame.  One galaxy that was observed 
on the problematic night, NGC 3384, has previously measured index profiles.  Applying 
the correction resulted in much better agreement with the published values (see 
Section 3.7).  No scattered light correction was applied on any of the other nights 
since the correction derived for those nights did not affect the line strength 
profiles significantly.

\subsection{Measuring and Correcting for Rotation and Velocity Dispersion}

We measured rotation and velocity dispersion in the stellar components of our 
galaxies using the pPXf package \citep{2004PASP..116..138C}.  SSP spectra 
were constructed using the SP models by \citet[hereafter 
BC03]{2003MNRAS.344.1000B} assuming a Chabrier initial mass function.  The pPXF 
routine fit each galactic extraction with a linear combination of SSP 
spectra, shifting and broadening these to match the galaxy's rotation and velocity 
dispersion respectively.  The fit was performed within the wavelength range
4800-5400 \mbox{\AA}, with the emission lines H$\beta$, [OIII] 4959, and [OIII] 
5007 masked out.  The profiles will be presented in Paper II.

The galaxy spectrum at each location was shifted by the measured stellar rotation 
before measuring the line indices.  To correct the indices for velocity dispersion, 
line strengths were measured on an SSP template that was broadened by the measured 
velocity dispersion and an unbroadened but otherwise identical template.  These 
templates were also constructed using a linear combination of BC03 SSP spectra with 
emission lines masked out, but the wavelength range for the fit was 4000-6600 
\mbox{\AA}, to include all the Lick indices.  This template was also used for 
emission correction (following section).  The measured absorption line equivalent 
widths were multiplied by the ratio of the unbroadened line strength to the broadened 
one; for magnitudes the correction factor is the difference between these two 
quantities.

\subsection{Measuring and Correcting for Emission}

Some absorption indices can be severely affected by line-filling by emission.  These
include the Balmer indices, Fe5015 (due to [OIII] 5007 emission), and Mg$_b$ 
(due to [NI] 5199 emission).  To correct for this, we subtracted the template 
described in the previous section from each galaxy spectrum.  If on the residual 
spectrum, H$\alpha$ was found to be in emission at the 5$\sigma$ level and a local 
maximum was detected at H$\beta$, a gaussian was fit to the H$\beta$ emission and 
subracted from the galaxy spectrum.  This procedure was repeated for H$\gamma$ and 
H$\delta$.  [OIII] 5007 and [NI] 5199 were subtracted out if they were found to be in 
emission at the 3$\sigma$ and 4$\sigma$ levels respectively.  A larger threshold was 
used for [NI] 5199 because this feature lies at the edge of the Mg$_2$ absorption 
feature and spurious discontinuities often show up there due to template mismatch.

EWs were measured for several emission-lines to study the nature of 
the ionized gas in bulges and inner disks.  The galaxy continuum, obtained by 
smoothing the galaxy spectrum, was added to the emission spectrum described above 
before measuring the EWs.  

%Table xxx gives the bandpasses used.

\subsection{Lick Index Measurements}

\begin{figure*}
\includegraphics[width=\textwidth]{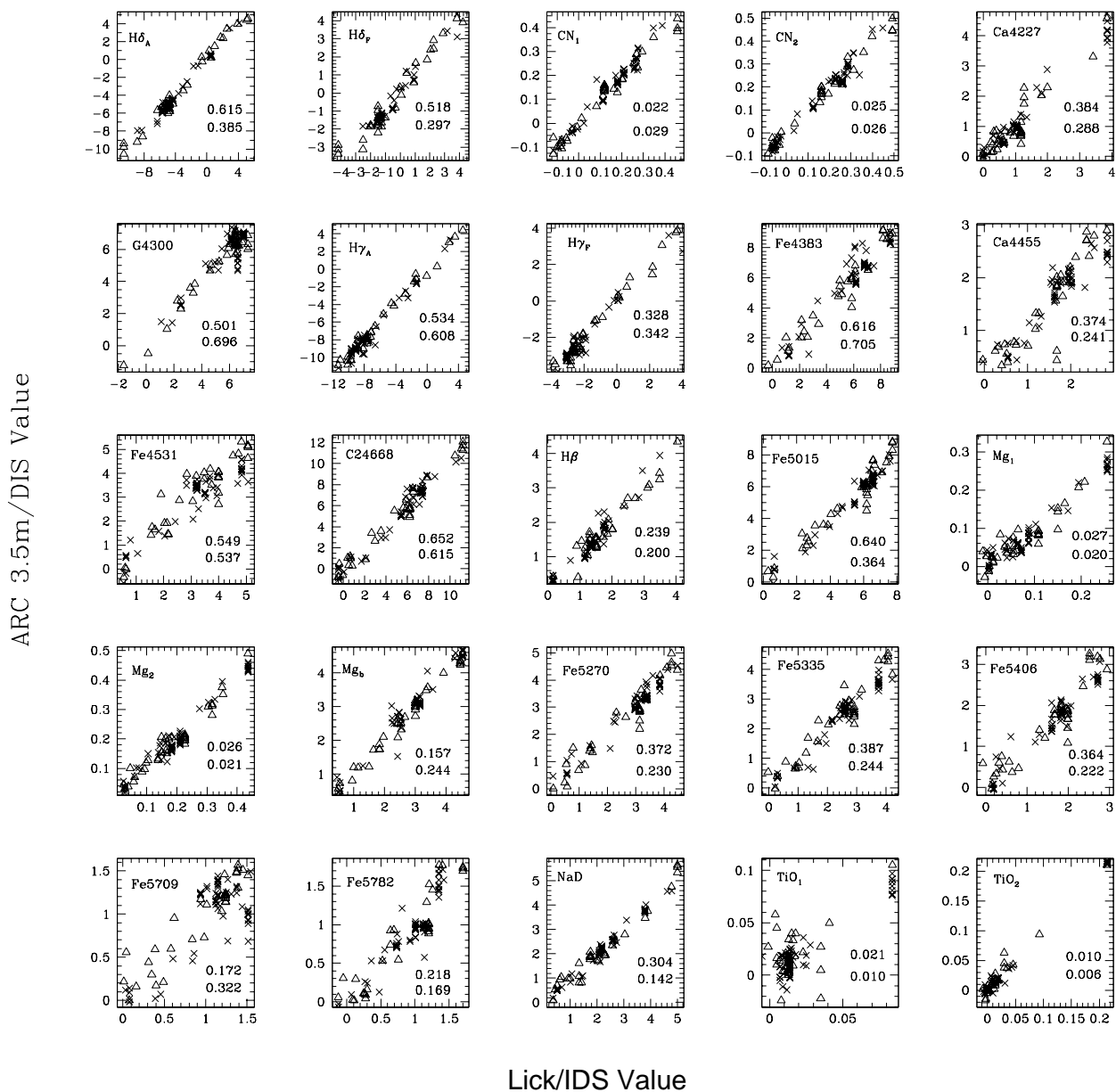}
\caption{Transformations to the Lick/IDS system.  Each point is a Lick standard star. 
 Triangles are DIS I.  Crosses are DIS III.  DIS III transformation were 
used for DIS II.  The rms scatter when applying the transformations on the stars are 
shown at the bottom-right of each plot (with DIS I above DIS III).}   
\label{trn}\end{figure*}

The final galaxy spectra were broadened to 9.5 \mbox{\AA} FWHM, which is 
approximately equal to the Lick resolution, and rebinned to a dispersion 0.125 
\AA/pixel.  Variable broadening as prescribed in \cite{1997ApJS..111..377W} was tried 
and found not to produce significantly different results.  The strengths of 25 
absorption features were measured using the latest bandpasses from Guy Worthey's 
webpage.  The EW or magnitude of each feature was computed following
\cite{1998ApJS..116....1T}.

Spectral indices were measured on Lick standard stars, exactly as done for the 
galaxies, to transform our line strengths to the Lick/IDS system.  Such 
transformations are necessary because our detector differs in resolution from the IDS 
and because the IDS spectra were not flux calibrated.  24 stars ranging in spectral 
type from F5V to K7III were used inderiving the transformations for DIS I and 22 
stars (also F5V to K7III) were used for DIS III.  DIS III transformations were used 
for DIS II since the same gratings were used for both.  From imaging one of the stars 
at several positions along the slit, it was determined that the line strengths do not 
vary significantly with slit position.  Fig. \ref{trn} shows the 
transformations.  The rms scatter when applying the transformations on the stars are 
shown at the bottom-right of each plot.  The flux response of the detector changes 
rapidly at the long wavelength end of the blue channel, making measurements there 
difficult.  Only the Mg$_1$ and Mg$_2$ indices were affected by this, but severely 
so, since their continuum bandpasses are far apart and the red continuum bandpass 
lies right where our flux response is steepest.  This resulted in large scatter in 
the transformations of these indices.  For the other indices, the scatter in the 
transformations is similar or slightly larger than that obtained by 
\cite{2002MNRAS.333..517P}.

\subsection{Comparison with the literature}

\begin{figure}
{\includegraphics[width =0.45\textwidth]{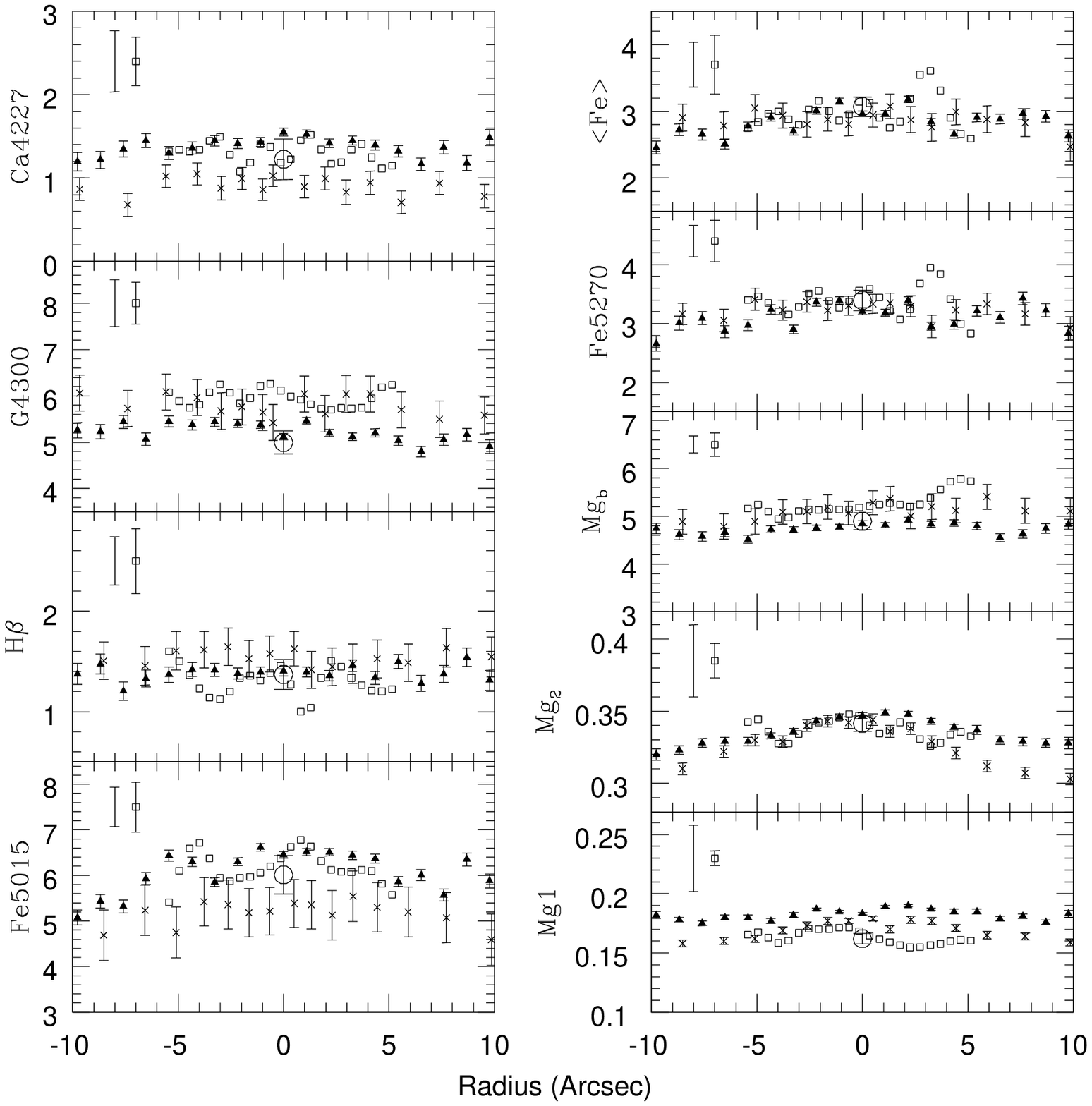}}
\caption{Comparison of line strength gradients between this work and 
published values for NGC 4472.  Triangles are from this work.  Large circles are 
from Trager et al. (1998).  Open squares are IFU data from Peletier et al. (1999) 
and crosses are long-slit data from Vazdekis et al.(1997).  At the top of each 
figure, error bars without points represent the scatter in stars from our 
transformations to the Lick system; error bars with squares indicate 
uncertainties in Peletier et al.'s data, including those involved in transforming to 
the Lick system.}\label{comp4472}
\end{figure}

Fig. \ref{comp4472} shows comparisons of Lick indices between this work and the 
most recently published values for NGC 4472.  There is excellent agreement in the 
Ca4227, H$\beta$, Fe5015, $\langle$Fe$\rangle$,and Fe5270 profiles between this work 
and all the published values.  Our other index profiles are systematically offset 
from at least one of the other two studies but the offset is within the uncertainties 
of transforming our data to the Lick system.  In general, our results agree better 
with the long-slit data from \cite{1997ApJS..111..203V} than the IFU data from
\cite{1999MNRAS.310..863P}.  Nearly all our central values agree with those 
of Trager et al.  Emission correction was not responsible for any disagreement among 
studies since NGC 4472 does not have much emission.

\begin{figure}
{\includegraphics[width =0.45\textwidth]{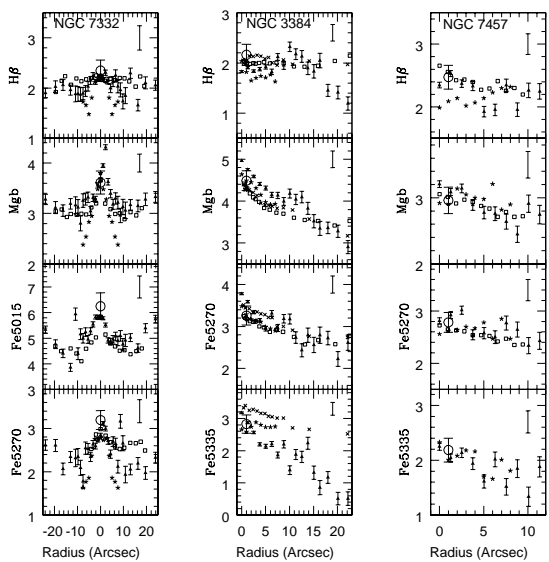}}
\caption{Comparison of line strength gradients between this work and published
values for NGCs 7332, 3384, and 7457.  Triangles are from this work.  Asterisks are 
azimuthally averaged IFU data from Sil'chenko (1999), Sil'chenko et al. (2003), and 
Sil'chenko et al. (2002) for NGCs 7332, 3384, and 7457 respectively.  Squares are 
SAURON data extracted along the major-axis from Falc{\' o}n-Barroso et al. (2004), de 
Zeeuw et al. (2002), and Sil'chenko et al. (2002).  For NGC 3384, crosses are 
long-slit data from Fisher et al. (1996).}
\label{compidx2}\end{figure}

The first column of Fig. \ref{compidx2} shows index comparisons between this 
work (triangles), azimuthally averaged IFU data from Sil'chenko (1999; asterisks), 
and SAURON data extracted along the major-axis from Falc{\' o}n-Barroso et al. (2004; 
squares) for NGC 7332.  All the central values are in agreement.  The slope of our 
H$\beta$ and Fe5270 profiles are steeper than Falc{\' o}n-Barroso et al.'s but not as 
steep as Silchenko's.  Our Mg$_b$ and Fe5015 profiles agree reasonably well with 
Falc{\' o}n-Barroso's but Silchenko's Mgb profile is again steeper.

The second column of Fig. \ref{compidx2} shows index comparisons between this work, 
long-slit data from Fisher et al. (1996; crosses), IFU data from Sil'chenko et al. 
(2003; asterisks), and SAURON data from de Zeeuw et al. (2002; squares) for NGC 3384. 
 Here also, Sil'chenko's profile is averaged azimuthally while de Zeeuw et al.'s is 
an extraction along the major axis.  This galaxy was observed on the night in which 
scattered light was corrected for as described in Section 3.2.  Fisher et al.'s 
profiles are in good agreement with those from SAURON.  Our Mg$_b$ and Fe5270 
profiles agree marginally with published values.  In the central regions, our 
H$\beta$ profile falls in between de Zeeuw et al.'s and Sil'chenko et al.'s.  Outside 
about 15 arcsec, our profile falls steeply, possibly due to inadequate scattered 
light correction, while de Zeeuw et al.'s stays flat.  Our Fe5335 profile is much 
steeper than the published ones especially outside 15 arcsec (again likely due to 
inadequate scattered light correction).  Two of the index-combinations studied in 
this paper, [MgFe]' and Mgb/$\langle Fe \rangle$, include Fe5335.  The discrepancy 
between our Fe5335 profile NGC 3384 and published ones does not affect [MgFe]' 
significantly.  Fisher et al.'s profile (cyan curve in Fig. \ref{met1}) shows good 
agreement with ours.  For Mgb/$\langle Fe\rangle$ however, we obtain a positive 
gradient while Fisher et al. found no gradient (Fig. \ref{alpha1}), possibly due to 
scattered light in our data.  The other two objects affected by scattered light are 
NGCs 2787 and 3945.  NGC 2787 shows no gradient in Mgb/$\langle Fe \rangle$.  NGC 
3945 has an asymmetric positive gradient.  This could not be due entirely to 
scattered light since the scattered profile was symmetric.

The third column of Fig. \ref{compidx2} compares profiles for NGC 7457 with  
IFU data from \cite{2002ApJ...577..668S} and archival SAURON data which were also 
presented in Silchenko et al.'s paper.  Silchenko et al.'s values are systematically 
smaller in the central regions than those obtained by others except for the case of 
Fe5335, where they are in agreement with ours.  Our values have larger scatter but 
are otherwise in agreement with the SAURON values.  The agreement among studies 
is worst for H$\beta$ although this galaxy has little or no Balmer emission.  The 2D 
SAURON map shows a negative gradient along the major axis but not along the minor 
axis while Sil'chenko et al.'s 2D map shows no gradients whatsoever; our values are 
in better agreement with SAURON's than Sil'chenko et al.'s.

\subsection{Absorption Line Indices and SSP Models}

We have measured all 25 Lick indices in our galaxies as a function of galactocentric 
radius.  In this paper, we concentrate on a subset of these indices which are 
sensitive to age, metallicity, and $\alpha$/Fe. The most age-sensitive indices are 
the Balmer indices H$\beta$, H$\gamma_A$, H$\gamma_F$, H$\delta_A$, and H$\delta_F$.  
Of these, H$\beta$ suffers most from line-filling by emission while the H$\delta$ 
indices are the least affected.  On the other hand, H$\beta$ offers the most 
orthogonality with respect to metallicity-sensitive indices.  Using a combination of 
the Balmer indices, we can obtain more reliable age estimates than with just one 
index.  For metal lines, we compute the indices Mgb/$\langle$Fe$\rangle$ and [MgFe]' 
as discussed in TMB; the former is directly related to $\alpha$/Fe, and [MgFe]' 
traces metallicity without any sensitivity to $\alpha$/Fe.  Individually, [MgFe]' and 
the Balmer indices are degenerate in age and metallicity but together they can break 
the degeneracy since each index has a different age-metallicity dependence.

The integrated-light spectrum of an object is a linear combination of SSPs.  Some 
objects, such as globular clusters, are well represented by a single SSP while 
galaxies are generally not.  Still, one can characterize the SPs of a 
galaxy by an ``equivalent SSP".  Since the integrated light from a galaxy is weighted 
by luminosity, its SSP age and metallicity is likely to be different from the 
mass-weighted age and metallicity of its stars.  SSP values are useful 
parameterizations of the SPs but cannot be interpreted as true ages 
and metallicities, since galaxies most likely contain a range of both.  To avoid 
over-interpreting our data, we focus on the line strengths, mentioning SSP values 
only to illustrate dramatic differences between objects or regions within an object 
(i.e. 2 vs 10 Gyr as opposed to 8 versus 12 Gyr).  Different line strengths in 
different objects (or within an object) imply different SPs; the SSP 
models allow us to infer the underlying source of the differences.

\subsection{Bulge-to-disk decomposition}

\begin{figure*}
\includegraphics[width=\textwidth]{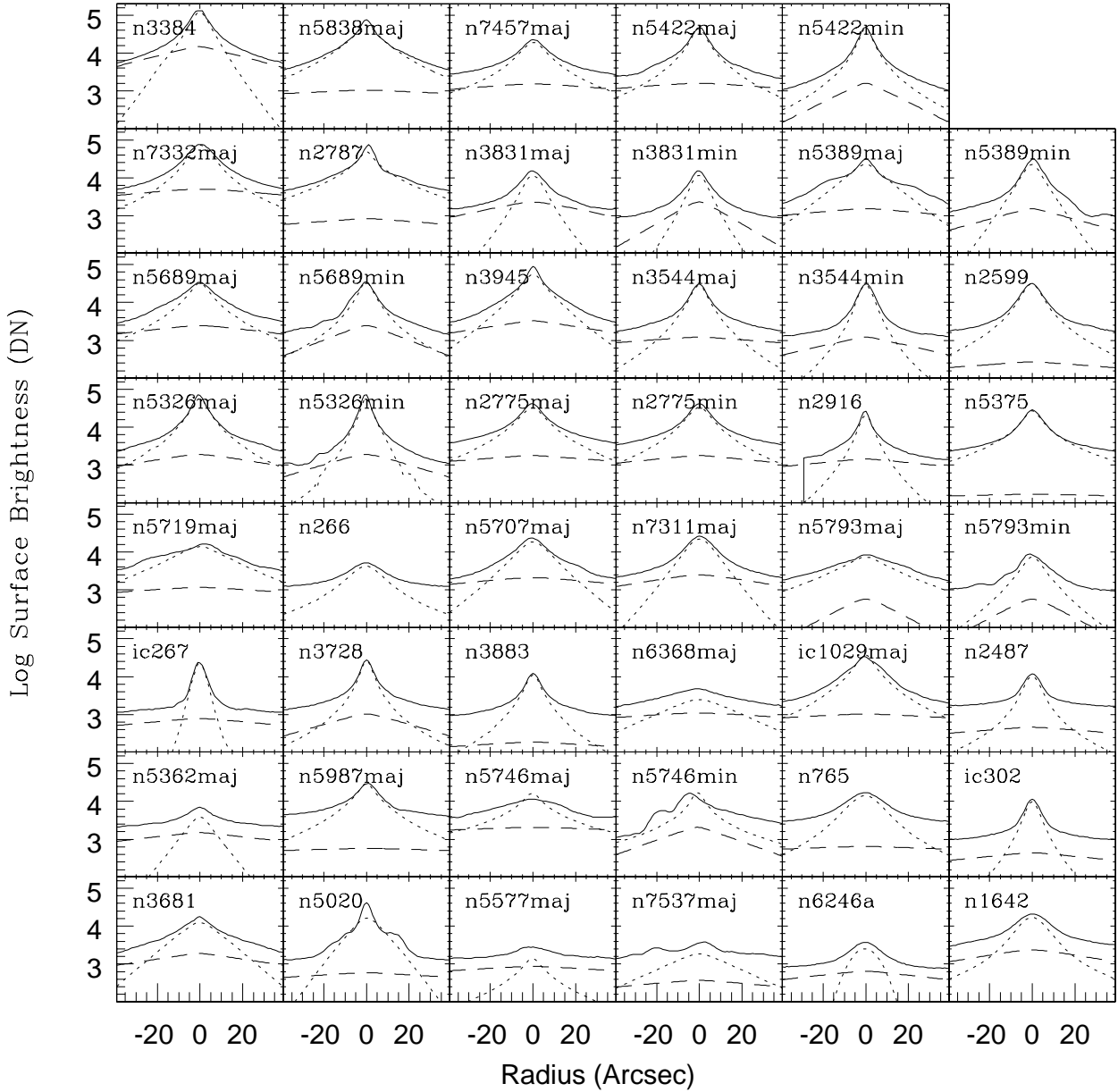}
\caption{Results of 2D bulge-to-disk decomposition.  The light profile of each galaxy 
is shown as a solid line, the best-fit bulge as a dotted line, and the best-fit disk 
as a dashed line.  The position angles are the same as those used for the 
spectroscopy (Table 3). }\label{bd}
\end{figure*}

One-dimensional and two-dimensional bulge-to-disk decomposition (B/D) was performed 
on our images.  The disk and bulge were simultaneously fit with an exponential and an
exp[r$^{1/n}$] profile, respectively.  Initial fits were made using fixed $n$ of 1, 
2, 3, and 4; the best fit of these was used as a starting guess in a final fit where 
$n$ was allowed to vary as a free parameter.  The results from the 1D decomposition 
were used as starting guesses for the 2D.  In the 2D fits, both the bulge and disk 
components were allowed to be elliptical.  This resulted in bars being fit as bulges 
in some galaxies (IC 302 and NGCs 266, 765, 3681, 3883, 5375) but not others (IC 267 
and NGCs 2487 and 5020).  A two-component bulge/disk model, such as ours, also 
overestimates the bulge component of our three S0s with luminous inner disks (NGCs 
2787, 3384, and 3945) as previously found by \cite{2003ApJ...597..929E}.  We pay 
close attention to these effects when studying the line strength profiles in the 
bulge- and disk-dominated regions (Section 4.2).  It was found through visual 
inspection that the best-fitting bulge and disk components of the other galaxies were 
reasonable. Fig. \ref{bd} shows the light profiles of our galaxies along with the 
best-fit bulge and disk components.

The bulge-to-disk decomposition used here primarily serves to determine the relative 
contribution of bulge and disk light as a function of radius;  while B/D 
decomposition is notoriously difficult, especially when allowing for a Sersic bulge, 
this ratio is determined more robustly than the derived values of bulge effective 
radius and Sersic index.

Throughout this paper, the term ``disk" refers to the exponential outer region of the 
luminosity profile and the term ``bulge" refers to the excess light in the inner 
regions that was fit with a Sersic profile.  It is possible that these 
photometrically-determined components do not 
correspond to a cold and a hot component, a flat and a spheroidal component, or a 
young and an old component.   Through simulations, \cite{2003ApJ...597...21A} found 
that while the stars of a simulated galaxy were well fit by a Sersic + exponential 
profile, the hot and cold components were not well fit individually by a Sersic and 
an exponential profile respectively.  Instead both were Sersic in the inner regions 
and exponential further out.  Using our kinematic information, we address this issue 
in Paper II.  The structural properties and how they relate to the SPs will also 
be presented in that paper. 
 
\section{Results}

\subsection{Central Line Strengths}

\begin{figure*}
\includegraphics[width=\textwidth]{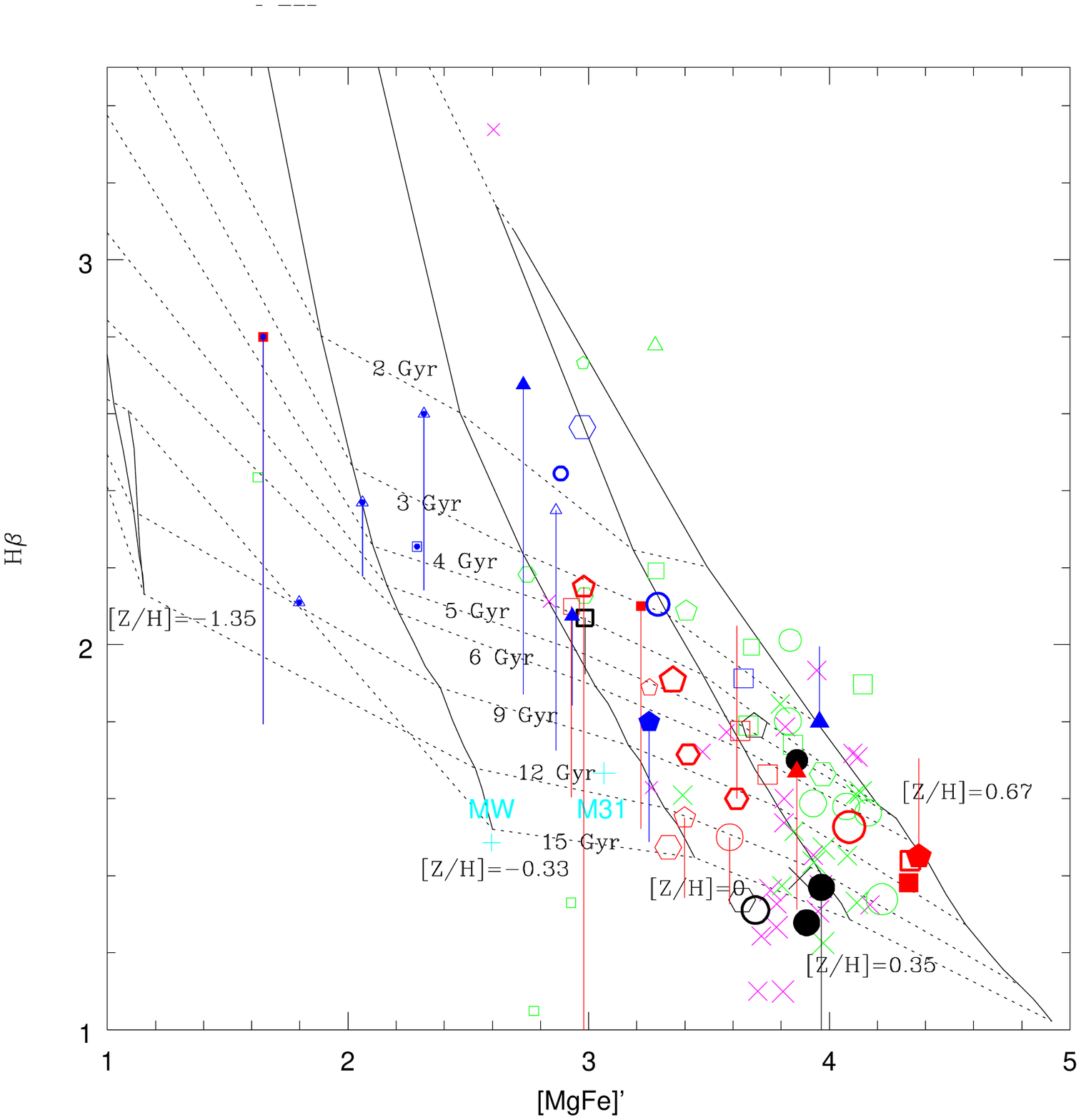}
\caption{H$\beta$ versus [MgFe]' in the central regions of bulges and ellipticals.  
Magenta crosses are galaxies from Trager et al. (1998) that were classified as 
elliptical and had H$\beta$ uncertainty less than 0.2 and velocity dispersion 
uncertainty less than or equal to 10 km s$^{-1}$.  Green symbols are bulges and 
ellipticals from Proctor \& Sansom (2002).  The Milky Way (Puzia et 
al. 2002) and M31 (Puzia et al. 2005) are shown as '+' symbols.  Symbols for our 
bulges and those of Proctor et al. are as in Fig. \ref{bksigvmax}.  Larger symbol 
sizes were used for objects with larger central velocity dispersion.  Bulges for 
which we do not have color information are in black.  The metal-poor (MP) bulges are 
marked with an additional blue dot.  Where spectra exist along both minor and major 
axes, the average of the two values was used.  If a point has an accompanying 
vertical line segment, its location was determined using an average of the other four 
Balmer indices, instead of H$\beta$, to compute the SSP age.  The other end of the 
line segment shows the H$\beta$ value.  TMB's models are superimposed on the 
plot.}\label{hbmetctr}\end{figure*}

Fig. \ref{hbmetctr} shows the measured central values of H$\beta$ and [MgFe]' for 
our sample as well as published values for the Milky Way 
\citep{2002A&A...395...45P}, M31 \citep{2005A&A...434..909P}, other bulges 
\citep{2002MNRAS.333..517P} and elliptical galaxies \citep{1998ApJS..116....1T, 
2002MNRAS.333..517P}.  Our central indices were measured on spectral 
extractions binned to match Trager et al.'s 4 arcsec aperture and Proctor et al.'s 
3.6 arcsec aperture as closely as possible (3.3, 4.2, and 3.8 arcsec for DIS I, II, 
and III respectively).  Symbols are as in Fig. \ref{bksigvmax}.  Larger symbols 
denote larger central velocity dispersions.  If a point has an accompanying vertical 
line segment, its location was determined using an average of the other four Balmer 
indices instead of H$\beta$ to compute the SSP age.  The other end of the line 
segment shows the H$\beta$ value.  If the difference between these is large, this is 
due most likely to errors in H$\beta$ emission correction since the galaxies whose 
H$\beta$ values lie outside the model grid have strong H$\beta$ emission.  Two 
galaxies have H$\beta$ values that put them outside the plot range.  These are NGCs 
2787 and 5719 with H$\beta$ values of -0.506 and 0.607 respectively.

Bulges can be divided into three groups according to which region of Fig. 
\ref{hbmetctr} they populate: the old metal-rich (OMR, lower right) region , the 
young metal-rich (YMR, upper right) region with ages less than 3 Gyr and super-solar 
metallicity, or the metal-poor (MP, left) region with sub-solar metallicity.  This 
classification scheme is analogous to that of Prugniel et al. with our MP, OMR, and 
YMR classes corresponding to their A, B, and C classes respectively.  Membership in a 
region is closely related to bulge color.  Most of the red bulges populate the OMR 
region while all but one of the bulges in the MP region are blue.  The exception is 
IC 267 (the red filled square with [MgFe]'=1.65).  This bulge has strong Balmer 
emission which suggests that its red color is due to dust from recent star formation 
as opposed to an old SP.  Hereafter, when we discuss blue bulges, we include IC 267 
since it is similar, in most respects, to blue bulges and since its B-K color puts it 
right on the borderline between blue and red classes anyway.  Color does not uniquely 
represent the SPs of blue bulges; some bulges are blue because they are metal-poor 
while others are blue because they are young.  To distinguish the two classes of blue 
bulges in subsequent plots, the MP bulges are marked with an additional blue dot.

Central line strengths are also sensitive to Hubble type.  All the early-type 
(S0-Sab) bulges have large central metallicities.  The blue early-types are in the 
YMR region while the red early-types are in the OMR region.  Bulges of late-type 
spirals are more heterogeneous in their central line strengths than those of 
early-types.  The MP region is populated exclusively by late-types but late-types are 
also found in the other two regions.  Trager et al.'s ellipticals mostly populate the 
OMR region with a couple of them lying in the YMR region.  Lower-luminosity 
ellipticals span a similar range of SSP values as our blue bulges 
\citep{2003AJ....125.2891C}; unfortunately, we cannot include Caldwell et al.'s data 
in Fig. \ref{hbmetctr} because they did not measure the same indices.  The bulges of 
the MW and M31 lie in the MP region.

\begin{figure*}
\includegraphics[width=\textwidth]{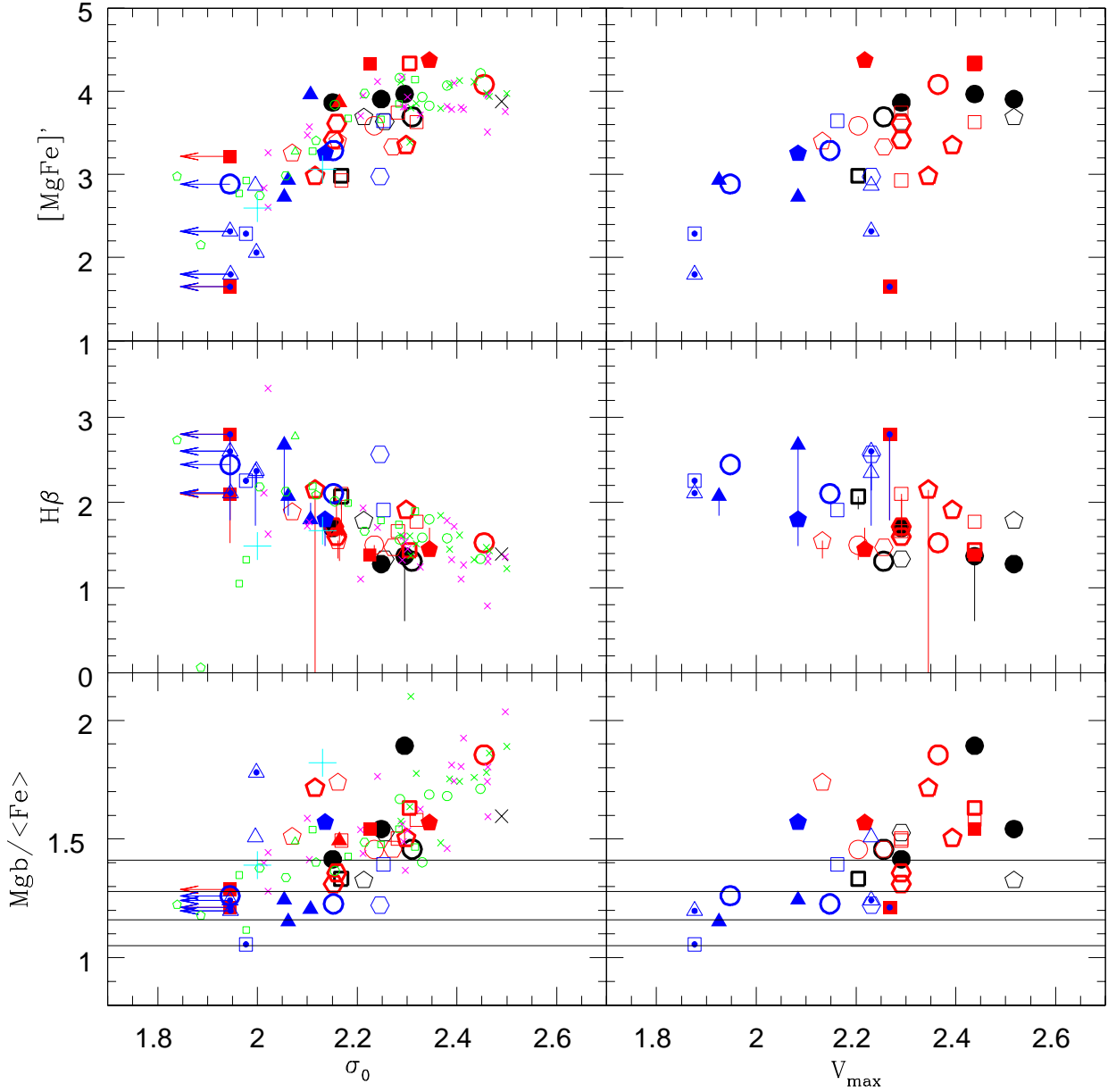}
\caption{Central line strengths versus central velocity dispersion and maximum disk 
rotational velocity.  Symbols as are in Fig. \ref{hbmetctr}.  Arrows represent 
upper limits for galaxies whose velocity dispersions are close to or below our 
resolution limit.  IC 302 and NGCs 765, 2487, 2916, and 6246A are not shown on the 
right panel since their inclinations are too low to measure rotation.  In the bottom 
panel, the region within the inner horizontal lines corresponds to models with 
solar $\alpha$/Fe for metallicities -1.35 $\leq$ [Z/H] $\leq$ 0.35 and ages from 8 to 
15 Gyr (from Fig. 4 of TMB).  The region within the outer horizontal lines represents 
models with the same metallicities and ages from 3 to 15 
Gyr.}\label{indexsigvmax}\end{figure*}

Much of the variation in central line strengths is due to correlations between the 
line strengths and the global kinematics.  The different types of bulges (MP, YMR, 
and OMR) and ellipticals form a continuous and overlapping sequence on a plot of 
[MgFe]' versus central velocity dispersion (Fig. \ref{indexsigvmax}, top-left).  In 
both bulges and ellipticals, [MgFe]' is correlated with $\sigma_0$ at the 
low-$\sigma_0$ end.  As $\sigma_0$ increases beyond $\log \sigma_0>$2.2, [MgFe]' 
remains constant.  Bulges show larger scatter than ellipticals in the 
[MgFe]'-$\sigma_0$ relation.

The apparent dichotomy seen in Fig. \ref{hbmetctr} between the MP bulges and the 
other objects (namely that the MP bulges have smaller central [MgFe]') is naturally 
explained by the wide range of $\sigma_0$ and V$_{max}$ values spanned by the 
different types of objects.  The MP bulges reside in considerably shallower potential 
wells than the bulges and ellipticals which populate the OMR and YMR regions.  It is 
important to note here that the predominant view that ellipticals are metal-rich 
\citep{2000AJ....119.1645T, 2003ApJ...585..694E, 2005MNRAS.356.1440D, 
2005ApJ...621..673T} applies only to high-$\sigma_0$ ellipticals.  All of Trager et 
al.'s ellipticals have $\sigma_0>100$ km s$^{-1}$.  The three ellipticals from Trager 
et al' with the smallest velocity dispersions ($100<\sigma_0<105$ km s$^{-1}$) 
include the two with the smallest SSP metallicity and the one with the smallest SSP 
age.  Caldwell et al. (2003) found that ellipticals with $\sigma_0<100$ km s$^{-1}$ 
span a similar range of SSP ages and metallicities as our blue bulges.

[MgFe]' is also correlated with the maximum disk rotational velocity (Fig. 
\ref{indexsigvmax}, top-right), as previously found by Prugniel et al.  The blue 
bulges with $\log V_{max}>2.2$ are significant outliers in the [MgFe]'-V$_{max}$, 
having smaller central values of [MgFe]' than their red counterparts.

Balmer indices are anti-correlated with $\sigma_0$ and weakly anti-correlated
with V$_{max}$ (middle panel of Fig. \ref{indexsigvmax}).  Residuals in the
H$\beta-\sigma_0$ and H$\beta-V_{max}$ relations are correlated with color such that
at a given $\sigma_0$ and V$_{max}$, blue bulges (both MP and YMR) have larger
H$\beta$ values than red bulges.

While individual indices are degenerate in age and metallicity, the different symbol
sizes in Fig. \ref{hbmetctr} allow us to determine how metallicity and age vary with 
$\sigma_0$.  Mean SSP metallicity and SSP age are larger in larger-$\sigma_0$ 
galaxies but at fixed $\sigma_0$ the scatter in SSP metallicity and SSP age are 
large.  At fixed $\sigma_0$, age and metallicity are known to be anti-correlated in 
ellipticals \citep{2000AJ....120..165T, 2002MNRAS.333..517P}.  The ellipticals 
from Trager et al. with large SSP age have smaller SSP metallicity than those with 
small SSP age.  Such an anticorrelation appears to exist for red bulges but with 
considerably larger scatter.  Interestingly, the small-$\sigma_0$ blue bulges also 
appear to have an age-metallicity anticorrelation.

In bulges and ellipticals, the $\alpha/Fe$ ratio as indicated by 
Mgb/$\langle$Fe$\rangle$, is correlated with $\sigma_0$ and V$_{max}$ (bottom panel 
of Fig. \ref{indexsigvmax}).  In these two plots, the region within the inner 
horizontal lines corresponds to models with solar $\alpha$/Fe for metallicities -1.35 
$\leq$ [Z/H] $\leq$ 0.35 and ages from 8 to 15 Gyr; the region within the outer lines 
represents models with the same metallicities and ages from 3 to 15 Gyr (from Fig. 4 
of TMB).  Red bulges and ellipticals show good overlap in the 
Mgb/$\langle$Fe$\rangle$-$\sigma_0$ diagram.  With a few exceptions, blue bulges are 
consistent with having solar $\alpha$/Fe.  Consequently, most blue bulges have 
smaller Mgb/$\langle$Fe$\rangle$ ratios than red bulges or ellipticals at a given 
value of $\sigma_0$ or V$_{max}$.  One of the blue bulges, NGC 6246A, has super-solar 
$\alpha/Fe$, large age, and small metallicity ([Z/H]$\tilde -0.8$) like MW halo 
stars.  The other two blue bulges with super-solar $\alpha/Fe$ have super-solar 
metallicity like the majority of red bulges.

\begin{figure*}
\includegraphics[width=\textwidth]{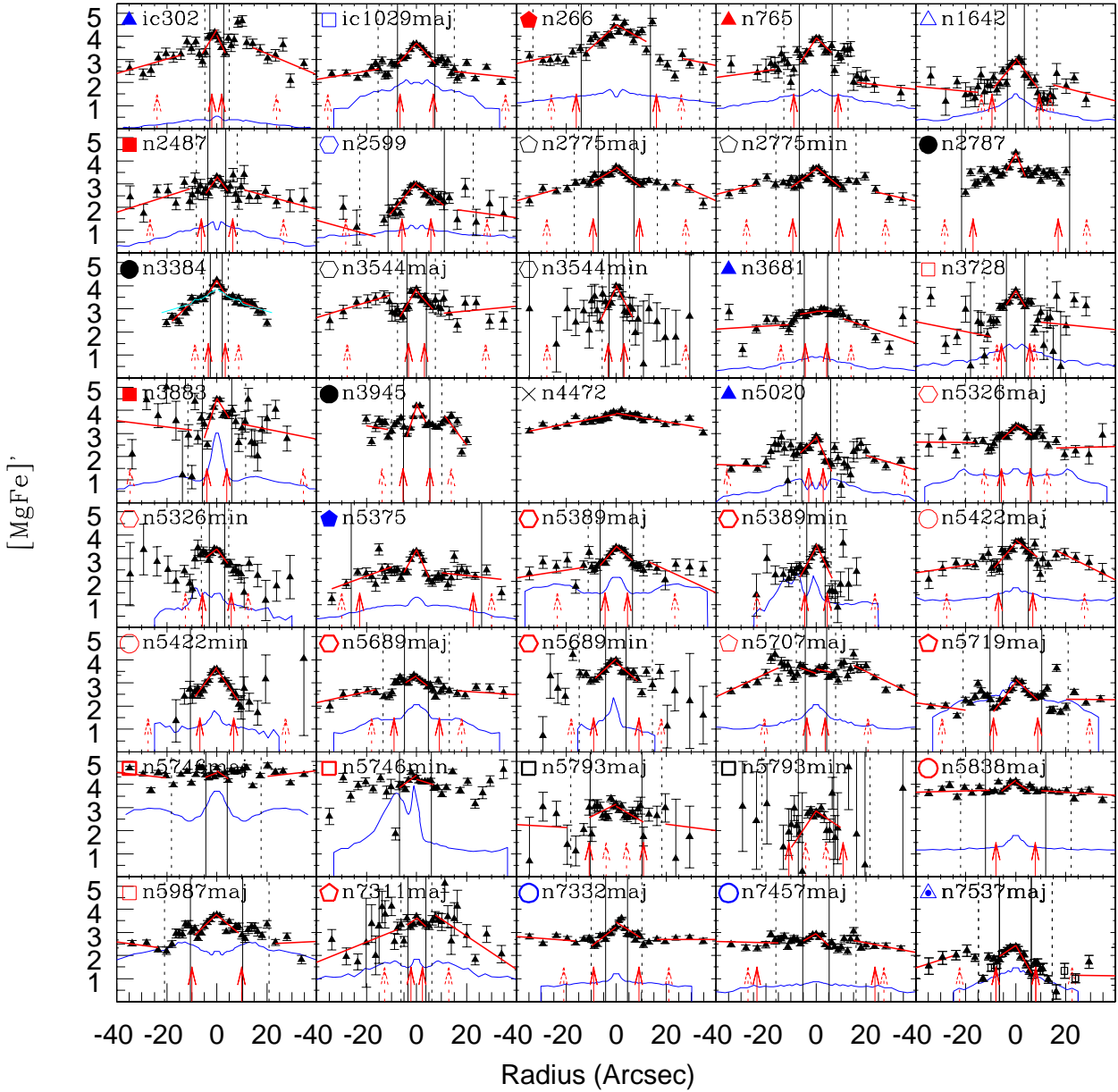}
\caption{[MgFe]' profiles in our large-bulge galaxies.  Open squares and filled 
triangles show points that were and were not corrected for [NI] 5199 emission 
respectively.  For NGC 3384, the cyan curve shows the profile obtained by 
Fisher et al. (1996).  At the top-left of each plot are symbols denoting the 
galaxy type (as in Fig. \ref{hbmetctr}) as well as the NGC or IC identifiers.  The 
solid and dotted vertical lines indicate where the ratio of bulge to disk light is 
two and half respectively, as determined from the bulge-to-disk decomposition.  The 
solid and dotted red arrows indicate the location of the bulge effective radius and 
disk scale length respectively.  Results from linear least-squares fits performed 
separately in the bulge- and disk-dominated regions are shown in red.  Blue lines are 
color profiles (B-K-3) from PB and DJ.  B-K was not available for IC 302; V-H-3 is 
shown instead.}\label{met1}\end{figure*}

There are hints that barred galaxies follow different index-$\sigma_0$ and 
index-V$_{max}$ relations than unbarred galaxies.  At fixed $\sigma_0$ and V$_{max}$, 
barred galaxies appear to have larger central values of [MgFe]' than unbarred 
galaxies (or galaxies with elliptical-shaped bulges) of the same $\sigma_0$ or 
V$_{max}$, with IC 267 being the most notable exception.  B/p bulges generally lie 
between and exhibit larger scatter than barred and unbarred/elliptical bulges in the 
[MgFe]'-$\sigma_0$ and [MgFe]'-V$_{max}$ diagrams.  The central regions of b/p bulges 
could be contaminated by the foreground disk resulting in smaller values of [MgFe]' 
than those of the low-inclination barred galaxies.  Barred galaxies appear to have 
smaller H$\beta$ than unbarred galaxies at fixed $\sigma_0$ but not at fixed 
V$_{max}$.  For $2.2>\sigma_0>2.35$, barred galaxies appear to have larger values of 
Mgb/$\langle$Fe$\rangle$ but the opposite is true for $2>\sigma_0>2.2$.  No striking 
difference is seen between barred and unbarred galaxies in the 
Mgb/$\langle$Fe$\rangle$-V$_{max}$ relation.  These results need to be verified using 
a large sample of barred and unbarred galaxies as a function of $\sigma_0$ and 
V$_{max}$.

The central regions of our three barred S0s with disk-like structure and kinematics 
(filled black circles) have super-solar SSP metallicities and $\alpha$/Fe ratios 
and two of the three have large SSP ages.  If all disks formed stars on long 
timescales (several Gyr), we would expect them to have small $\alpha$/Fe ratios, like 
the MW disk at the solar neighborhood, since ISM enrichment would eventually be 
dominated by SN Ia.  The ``luminous inner disks" of these S0s have not had such a 
star-formation history.  \cite{1999MNRAS.310..863P} found that the Sombrero galaxy is 
also dominated by a fast rotating disk whose [Mg/Fe] is similar not to other disks 
but to ellipticals of similar mass as the Sombrero.  

\begin{figure}
{\includegraphics[width =0.45\textwidth]{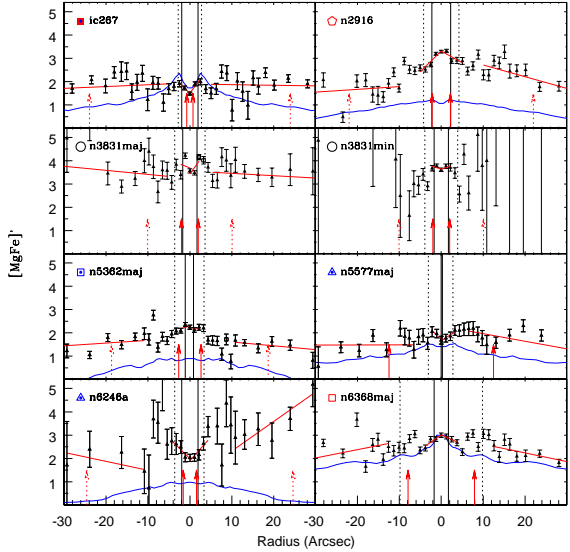}}
\caption{[MgFe]' profiles in our small-bulge galaxies.  Symbols are as in 
Fig. \ref{hbmetctr}.}\label{met2}
\end{figure}

\begin{figure}
{\includegraphics[width =0.45\textwidth]{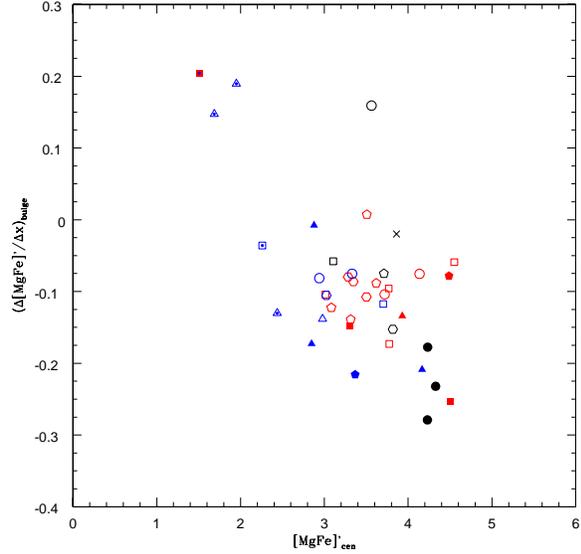}}
\caption{The [MgFe]' gradient within the bulge-dominated region versus the central 
value of [MgFe]'.  Symbols are as in Fig. \ref{hbmetctr}.}\label{bgrad}
\end{figure}

\begin{figure}
{\includegraphics[width =0.45\textwidth]{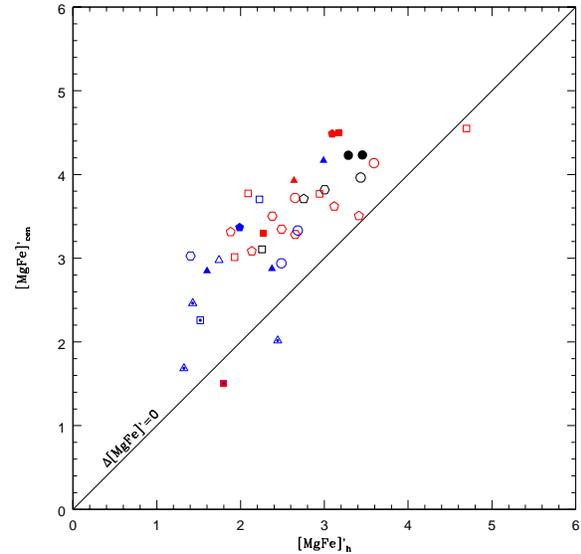}}
\caption{The central value of [MgFe]' versus the value at one disk scale length 
computed using the results from a least-squares fits to the [MgFe]' profiles.  The 
solid line shows where objects would lie if their bulges and disks were identical in 
[MgFe]'.  Symbols are as in Fig. \ref{hbmetctr}.}\label{bdmgfe}\end{figure}

\begin{figure}
{\includegraphics[width =0.45\textwidth]{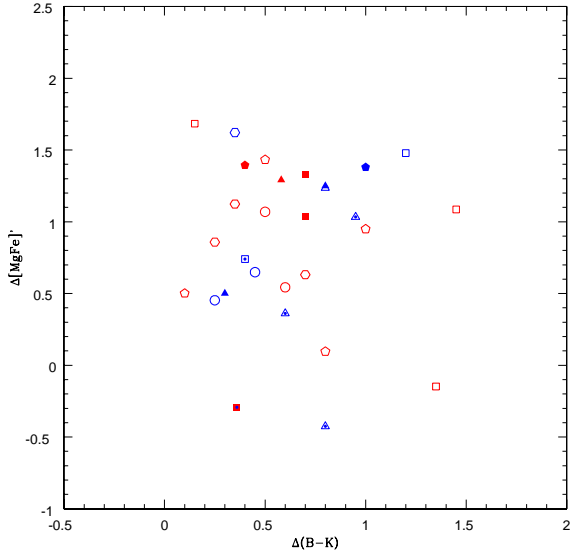}}
\caption{[MgFe]' gradient versus B-K color gradient, where a gradient is defined as 
the difference between the value at the center of the galaxy and the value at the 
disk scale length.  Symbols are as in Fig. \ref{hbmetctr}.}\label{mgfecol}
\end{figure}

\subsection{Line Strength Gradients}

\begin{figure*}
\includegraphics[width=\textwidth]{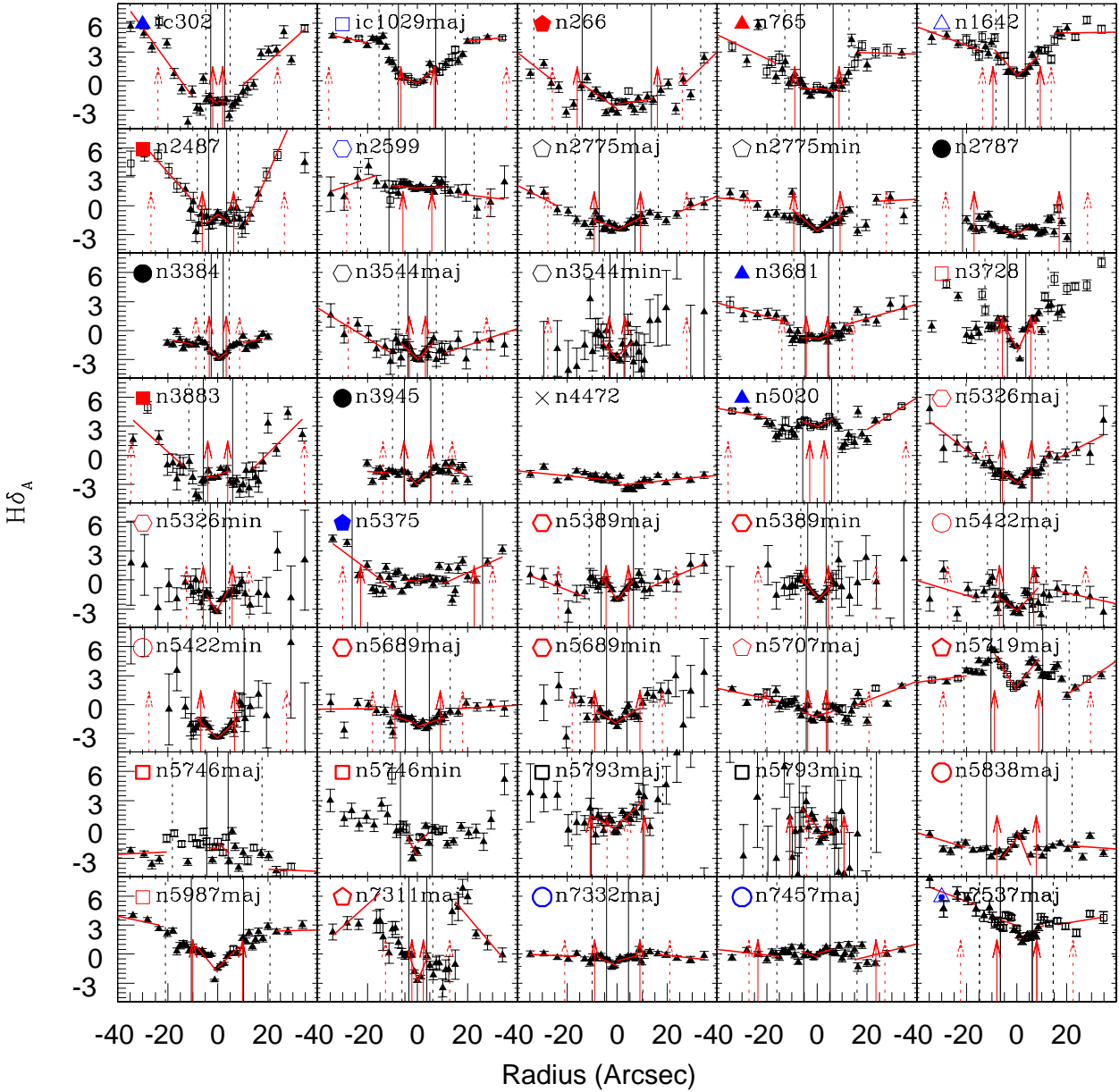}
\caption{H$\delta_A$ profiles in our large-bulge galaxies.  Symbols are as in Fig. 
\ref{hbmetctr}.}\label{hda1}
\end{figure*}

Figs. \ref{met1}-\ref{met2} show spatially resolved [MgFe]' profiles in our galaxies. 
They are split into two figures according to whether they have large or small 
bulges.  Open squares and filled triangles show points that were and were not 
corrected for [NI] 5199 emission respectively; the correction was seldom performed.
The solid and dotted vertical lines indicate where the ratio of bulge to disk light 
is two and one half respectively, as determined from the bulge-to-disk decomposition. 
 In the central regions, indices were measured on approximately 1 arcsec bins.

The large-bulge galaxies have five or more measurements within the solid vertical 
lines while the small bulge galaxies have three or fewer.  We performed linear 
least-squares fits to the [MgFe]' profiles separately in the bulge- and 
disk-dominated regions, selecting the points to include based on the bulge-to-disk 
decomposition but making exceptions where they seemed appropriate (such as when the 
bar was fit as a bulge; see Section 3.8).  We usually avoided the transition region 
from bulge to disk dominance.  The best-fit lines are shown in red.  
The large-bulge galaxies have negative [MgFe]' gradients (weaker [MgFe]' with 
increasing radius) in the bulge-dominated region.  In all but two of these, [MgFe]' 
decreases steadily from the galaxy center to the solid vertical lines, beyond which 
the slope of the profile changes.  The exceptions are NGCs 3681 and 5707, 
where [MgFe]' is constant within the solid vertical lines but decreases in the 
bulge-disk transition.  In low-inclination galaxies and along the major axes of 
inclined galaxies, [MgFe]' is usually larger just outside the solid vertical lines 
than inside it (NGC 5987 is a good example).  The fact that we can identify distinct 
bulge and disk components in the [MgFe]' profiles suggests that disk contamination 
is not significant within the solid vertical lines.

Another test for disk contamination is how the major and minor axis profiles vary as 
a function of inclination.  Low-inclination galaxies should have identical profiles 
if there are no azimuthal differences in line strengths.  The major and minor axis 
profiles of NGC 2775, the only low-inclination galaxy for which both were obtained, 
are indeed identical.  The minor axis profiles of the edge-on large-bulge 
galaxies (NGCs 5422, 5689, 5746, 5793) continue decreasing beyond the distance at 
which the major axis profiles flatten off, a result previously obtained by 
\cite{1996ApJ...459..110F} for edge-on S0s.  This implies that the outer bulge has 
lower metallicity than the inner disk.  At intermediate inclinations, one expects 
more disk contamination on one side of the minor axis (the dusty side) than the major 
axis for the same solid angle.  In thethree intermediate inclination galaxies for 
which we have major and minor axis spectra (NGCs 3544, 5326, and 5389), this effect 
is clearly seen; the profile is asymmetric outside the solid vertical lines with the 
profile turning over at a smaller galactocentric distance on the dusty (left) side 
than on the dust-free side.  Within the solid lines, there is good agreement in 
gradient slopes between major and minor axes, indicating that disk contamination is 
not significant.

The small-bulge galaxies generally do not have negative [MgFe]' gradients within the
bulge-dominated region.  In four of them (IC 267 and NGCs 3831, 5577, and 6246A), 
there is the hint of a positive gradient while the rest (NGCs 2916, 5362, and 6368) 
are consistent with having little or no gradient.  Since the profiles are always 
different inside and outside the solid lines, they cannot be explained by disk 
contamination.

The slopes of the [MgFe]' gradients within the bulge are shown in Fig. \ref{bgrad} as 
a function of the central [MgFe]' values.  Galaxies with large central values have 
correspondingly large negative gradients while the three galaxies with the smallest 
central values, namely IC 267 and NGCs 5577 and 6246A, have positive gradients.  
Since central [MgFe]' is correlated with $\sigma_0$ and V$_{max}$, 
it follows that the slope of [MgFe]' gradients in bulges is correlated with the 
global kinematics.

Most galaxies also have negative gradients in the disk-dominated region but it is 
shallower than that of the bulge.  Some galaxies (e.g. NGCs 5746, 5838, and 7332] 
have no gradient in the disk.

The [MgFe]' value at one disk scale length (computed using the results of our 
least-squares fits) is correlated with the central value (Fig. \ref{bdmgfe}).  This 
indicates that the metallicity of the disk is correlated with that of the bulge.  
This correlation holds for all galaxies, not just those with bars, blue bulges, or 
bulges identified as having disk-like structural or kinematical properties.

The blue lines in Figs. \ref{met1}-\ref{met2} are color profiles ($B-K-2$) from PB 
and DJ.  The shapes of the [MgFe]' and color profiles agree often but not always. 
Discrepancies occur most often in the central regions.  Several 
galaxies (e.g. IC 267 and NGC 266, 2487, 2916, 3728, and 5577) show a positive color 
gradient in the central 5 arcsec; of these, only IC 267 and NGC 5577 have a 
corresponding positive [MgFe]' gradient.  On the other hand, the central region of 
NGC 6246A has a positive [MgFe]' gradient but its color profile is flat.  de Jong 
(1996) noted that it is not possible to identify distinct bulge and disk components 
using the color profiles. However, it is possible to do so using the [MgFe]' 
profiles.  As mentioned earlier, the slopes of the [MgFe]' profiles are almost 
always distinct inside and outside the bulge-dominated region, with the former 
usually having a steeper negative gradient.  11 out of the 14 DJ galaxies and 15 out 
of the 17 PB galaxies have negative [MgFe]' gradients in the bulge-dominated region.  
The majority of the PB galaxies (12 out of 17 as opposed to 4 out of the 14 DJ 
galaxies) also show a negative color gradient in the bulge-dominated region.  The 
systematic difference in color gradients between the PB and DJ samples is likely due 
to different amounts of extinction at different inclinations since the PB galaxies 
have large inclinations while the DJ galaxies have small ones.

If we compare the gradient in [MgFe]' from the galaxy center to the disk scale length 
(computed using the fit results) with the color gradient (read from the profiles), we 
find that these two quantities are not tightly correlated (Fig. \ref{mgfecol}).  This 
is most likely due to the discrepancy between [MgFe]' and color profiles in the 
central regions of some galaxies, which in turn is due most likely to the color 
profiles being affected by dust.  

\begin{figure}
{\includegraphics[width =0.45\textwidth]{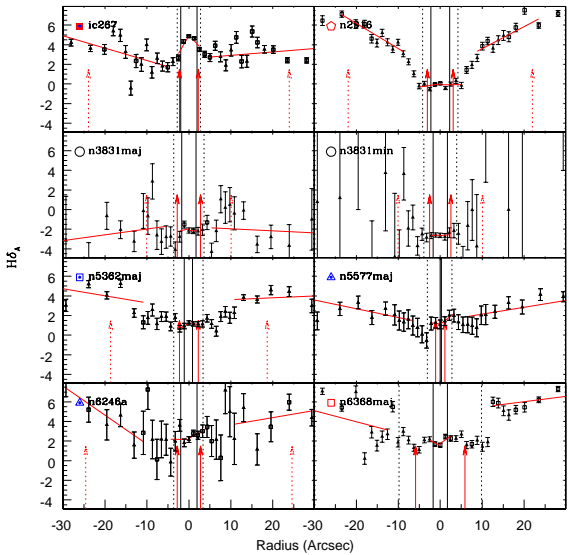}}
\caption{H$\delta_A$ profiles in our small-bulge galaxies.  Symbols are as in 
Fig. \ref{hbmetctr}.}\label{hda2}
\end{figure}

\begin{figure*}
\includegraphics[width=\textwidth]{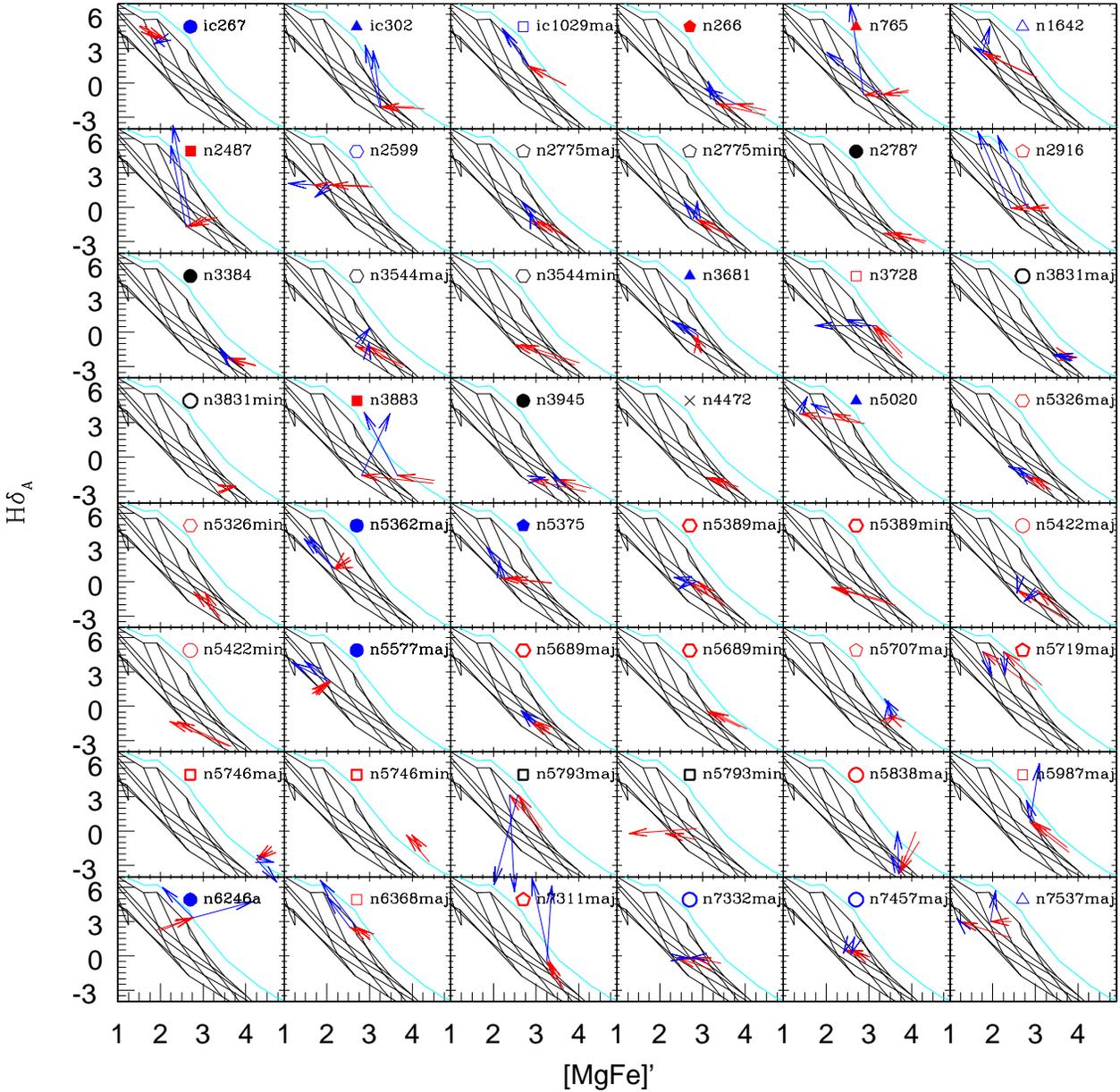}
\caption{Gradients in H$\delta_A$ and [MgFe]'.  Red arrows are drawn from the 
galaxy center to the edge of the bulge-dominated region on either side of the galaxy. 
 Blue arrows are drawn from the edge of the bulge-dominated region to the disk scale 
length.  Symbols are as in Fig. \ref{hbmetctr}.  TMB models with solar $\alpha$/Fe, 
age=1, 2, 3, 6, and 15 Gyr, and [Z/H]=-1.35, -0.33, 0, 0.35, and 0.67 are 
overlaid.  Unlike H$\beta$, the higher order Balmer indices are not independent of 
$\alpha$/Fe.  $\alpha$-enhanced models are parallel to the solar models, lying above 
and to the right.  The cyan curves show models with age=1 Gyr and [Z/H]=0.67 for 
[$\alpha/Fe$]=0.3.}\label{hdamet}\end{figure*}

\begin{figure*}
\includegraphics[width=\textwidth]{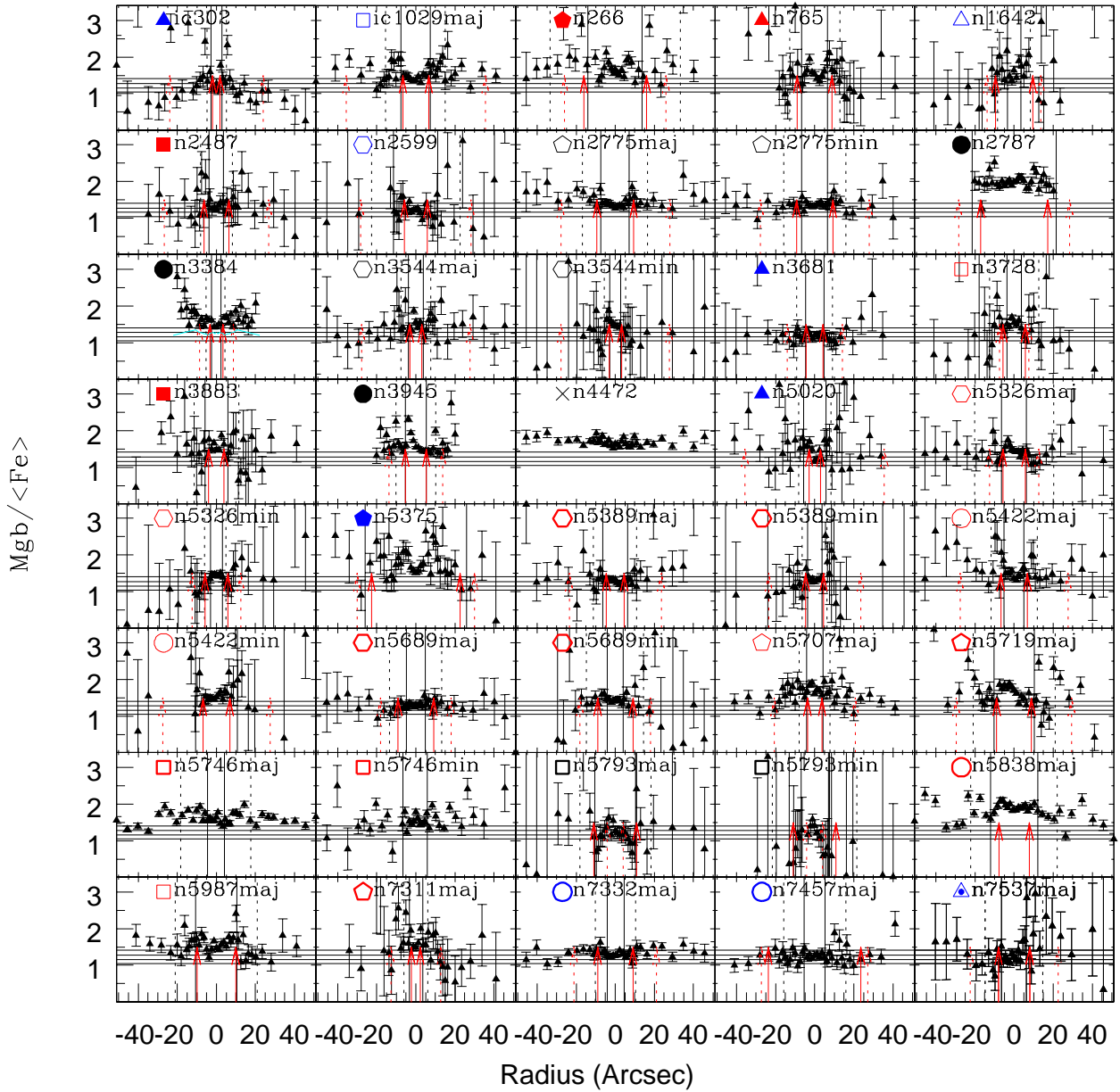}
\caption{Mgb/$\langle Fe \rangle$ profiles in our small-bulge galaxies.  For NGC 
3384, the cyan curve shows the profile obtained by Fisher et al. 
(1996).  Symbols are as in Fig. \ref{hbmetctr}.  The 
horizontal lines are as in \ref{indexsigvmax}.}\label{alpha1}\end{figure*}

\begin{figure}
{\includegraphics[width =0.45\textwidth]{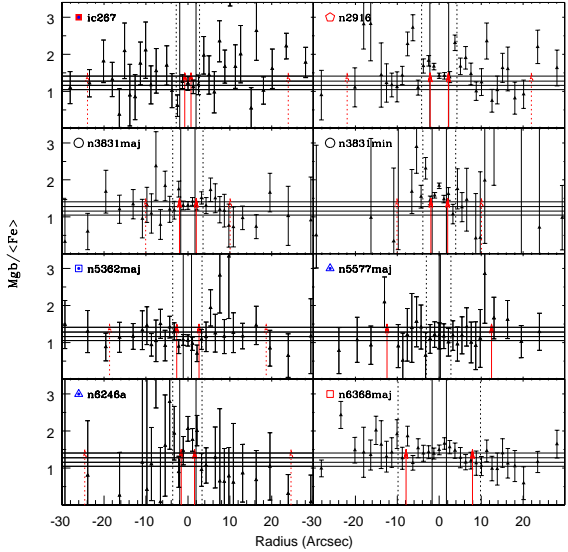}}
\caption{Mgb/$\langle Fe \rangle$ profiles in our small-bulge galaxies.  Symbols are 
as in Fig. \ref{hbmetctr}. The horizontal lines are as in 
\ref{indexsigvmax}.}\label{alpha2}\end{figure}

\begin{figure}
{\includegraphics[width =0.45\textwidth]{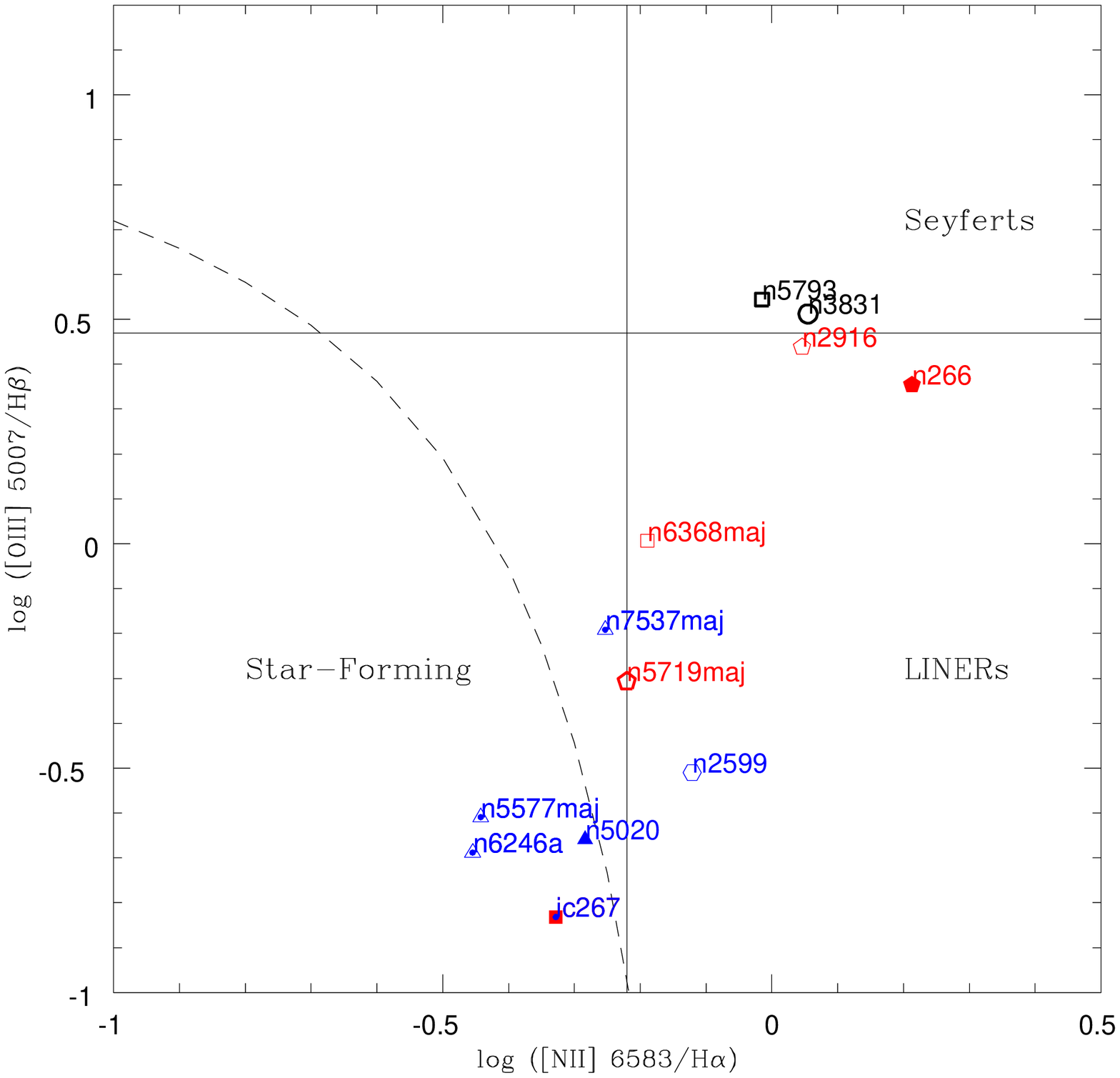}}
\caption{Bulges on the BPT (Baldwin et al. 1981) diagram in which the emission 
line flux ratio $[OIII] 5007/H\beta$ is plotted against the ratio $[NII] 
6583/H\alpha$.  Symbols are as in Fig. \ref{met1}. The dashed curve shows the 
demarkation between starburst galaxies and AGN as defined by 
Kauffmann et al. (2003).  We adopt the commonly used definition of LINERs having 
[OIII] 5007/H$\beta <$ 3 and[NII] 6583/H$\alpha >$ 0.6 and Seyferts having [OIII] 
5007/H$\beta >$ 3 and[NII] 6583/H$\alpha >$ 0.6.}\label{emissctr}
\end{figure}

\cite{2001AJ....122.1298G} found a greater prevalence of null or positive color 
gradients in barred galaxies than in unbarred galaxies.  They interpreted their 
result as evidence for gradients being erased by bar-driven mixing.  We do not see 
any systematic difference between barred and unbarred galaxies with regard to their 
gradients.  Also, if bars homogenized the SPs, we might expect a smooth transition in 
the line strength profiles from the bulge to the bar.  However, the outer bulges of 
barred galaxies have lower [MgFe]' than the inner bar the same way the outer bulges 
of unbarred galaxies have have lower [MgFe]' than the inner disk.

\subsubsection{Separating Age and Metallicity Effects}

Fig. \ref{hda1} shows gradients in the H$\delta_A$ index.  The H$\delta$ 
profiles show less scatter than the lower-order Balmer indices since they are less 
affected by emission.  Squares and filled triangles show points that were and were 
not corrected for emission, respectively.  Least-squares fits were performed on the 
Balmer indices exactly in the same manner as for the [MgFe]' profiles.  The fit 
results for H$\delta_A$ are shown in red in Fig. \ref{hda1}.  Gradient slopes 
computed on individual indices were combined to disentangle the effects of age and 
metallicity.   This is shown in Fig. \ref{hdamet} for the [MgFe]'-H$\delta_A$ index 
combination.  A red arrow is drawn from the galaxy center to the edge of the 
bulge-dominated region (on either side of the galaxy) and a blue arrow is drawn from 
there to the disk scale length.

The majority of galaxies (at least 29 out of 38) have negative metallicity 
gradients in the bulge-dominated region.  NGCs 3681, 3831, 5362, 5707, and 7311 show 
little or no metallicity gradient in the bulge.  Three of the five MP bulges (IC 267 
and NGC 5577 and 6246A) have positive metallicity gradients.  The remaining two 
galaxies (NGCs 5746 and 5793, both edge-on with b/p bulges) show internal 
discrepancies in the fit results.  Except for its minor axis H$\delta_A$ profile, NGC 
5746 is consistent with having little or no metallicity gradient. The minor axis 
profiles of NGC 5793 consistently a negative metallicity gradient but the major axis 
profiles show none.

The majority of galaxies are consistent with having little or no age gradient within 
the bulge.  At least ten galaxies (IC 302 and NGCs 266, 765, 2487, 2599, 2916, 3883, 
5020, 5375, and 5838) have a positive age gradient (larger age with increasing 
radius).  Of these, seven are barred and one has a b/p bulge.  At least five galaxies 
(NGCs 3728, 5577, 6246A, 7311, and the major axis of NGC 5793) have a negative age 
gradient in the bulge.

The Balmer indices show large scatter in the disk-dominated regions, making it 
difficult to determine whether metallicity and age gradients are present.  Plots 
of gradients in the five Balmer indices versus [MgFe]' (such as Fig. \ref{hdamet}) do 
not show consistent results for approximately half the galaxies.  Of the 
other half, most have a negative age gradient and little or no metallicity gradient.  
Most disks are solar or sub-solar in metallicity but NGCs 5746 and 5838 are 
super-solar well into the disk-dominated region.

\subsubsection{Abundance Ratio Gradients}

Most galaxies either have a positive gradient or no gradient in Mgb/$\langle Fe 
\rangle$ within the bulge-dominated region (Figs. \ref{alpha1} and \ref{alpha2}).  
The disk-dominated regions generally have solar $\alpha$/Fe.  Since the red-bulge 
galaxies have super-solar $\alpha$/Fe in the center, they have a negative gradient in 
the bulge-disk transition.  The blue-bulge galaxies have solar $\alpha$/Fe in 
the center.  These either have uniformly solar $\alpha$/Fe or a positive gradient in 
the bulge and a negative gradient in the bulge-disk transition.  Recall that what is 
marked as the bulge-dominated region in some galaxies (e.g. NGCs 266 and 5375) is 
actually a bar and that the true bulge-dominated region is smaller.  The elliptical 
galaxy, NGC 4472, is uniformly super-solar in $\alpha$/Fe.

There are a few galaxies that have super-solar $\alpha$/Fe in the disk-dominated 
region.  NGCs 266, 5707, and 5746 are nearly uniformly super-solar.  The disk of NGC 
5838 is super-solar but less enhanced than its bulge.

\subsection{Emission Lines in Bulges}

\begin{figure*}
\includegraphics[width=\textwidth]{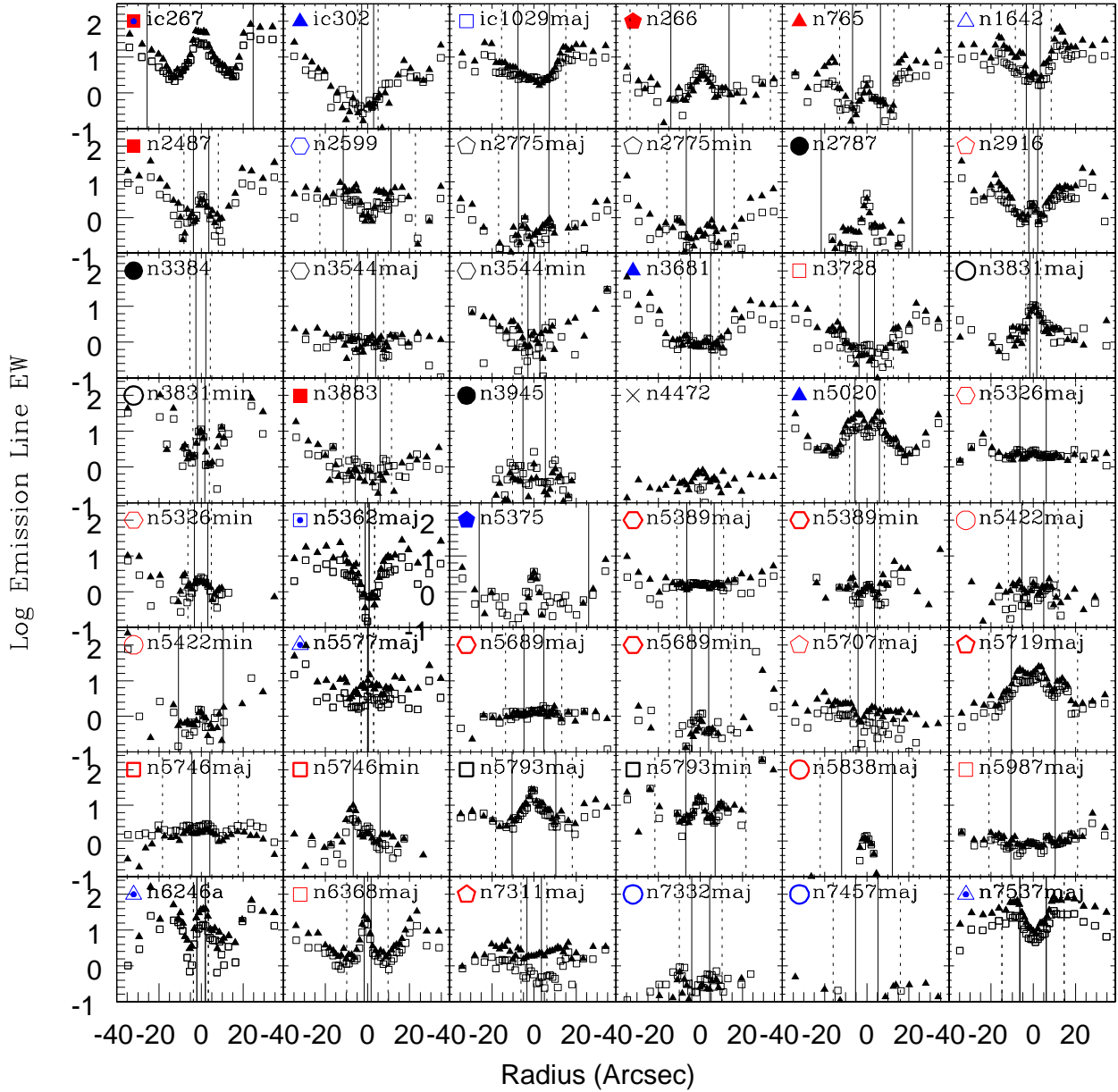}
\caption{Profiles of emission line strengths in bulges.  Triangles 
are H$\alpha$.  Squares are [NII] 6583.  Symbols are as in Fig. 
\ref{hbmetctr}.}\label{emiss}\end{figure*}

Fig. \ref{emiss} shows profiles of H$\alpha$ and [NII] 6583 emission strength in our 
galaxies.  We detect emission in the central regions of all our galaxies 
except the S0s NGCs 3384 and 7457.  The locations of our galaxies on the BPT 
\citep{1981PASP...93....5B} diagram of emission line ratios is shown in Fig. 
\ref{emissctr} for objects with central emission-line EW smaller than -0.5 \mbox{\AA} 
(the negative sign denotes emission) in H$\alpha$, [NII] 6583, H$\beta$, and [OIII] 
5007 .  The dashed curve shows the demarkation between starburst galaxies and AGN as 
defined by \cite{2003MNRAS.346.1055K}.  The majority of our emission-line galaxies 
are AGN.  If the emission in these galaxies is due entirely to the AGN, we would 
expect it to be restricted to the center of the galaxy.  However, in the three of 
these (NGCs 2599, 5719 and 7537), it is not centrally peaked.  In the other five 
(NGCs 266, 2916, 3831, 5793, and 6368), the emission peaks at the center and 
decreases steadily out to the edge of the bulge-dominated region, beyond which it 
rises again.  Therefore, all or most of the AGN also have active star formation in 
the bulge-dominated region.

AGN have previously been found to have a larger fraction of young stars than 
quiescent galaxies \citep{2001MNRAS.324.1087R, 2003MNRAS.339..772R}.  In agreement 
with these results, we find that most of the AGN have small SSP ages ($<4$ Gyr).  The 
only one with a large SSP age ($\tilde 15$ Gyr) is NGC 3831.  This could be due to 
errors from emission correction or from the young component not dominating the total 
luminosity.  \cite{2001A&A...366...68P} found that bulges with emission were small 
and metal-poor.  The star-forming region of our BPT diagram is populated by four blue 
bulges.  They have similar stellar populations as Prugniel et al.'s emission-line 
galaxies except that one of them (NGC 6246A) has a large SSP age, again possibly due 
to errors in emission correction.  

We see a wide range of behaviors in the emission-line profiles.  In some galaxies, 
such as NGCs 3681 and 5362, there is strong emission in the disk-dominated region but 
little or no emission in the bulge-dominated region as would be expected if disks 
continue to form stars while bulges do not.  The only galaxies with little or no 
emission in the disk-dominated region are S0s.  In other cases, there is emission 
throughout the galaxy but it is weaker in the bulge-dominated region (e.g. NGCs 1642, 
2916, 5020, and 7537).  This is consistent with a quiescent bulge and a star-forming 
disk coexisting in the central regions with the ratio of bulge to disk dominance 
decreasing with radius.  Alternatively, the bulge and disk could both be forming 
stars but the disk more actively so.  Finally, in some cases (e.g. IC 267 and NGCs 
266 and 5793), the emission lines are strongest at the center.

\section{The Formation of Bulges}

As mentioned in the introduction, present-day $\Lambda$CDM cosmology argues against 
the monolithic collapse scenario as does observational evidence for the recent and 
continuing mass assembly of ellipticals.  Of the main proposed formation scenarios, 
that leaves mergers and secular evolution as possibilities for bulges.  

\begin{figure*}
\includegraphics[width=\textwidth]{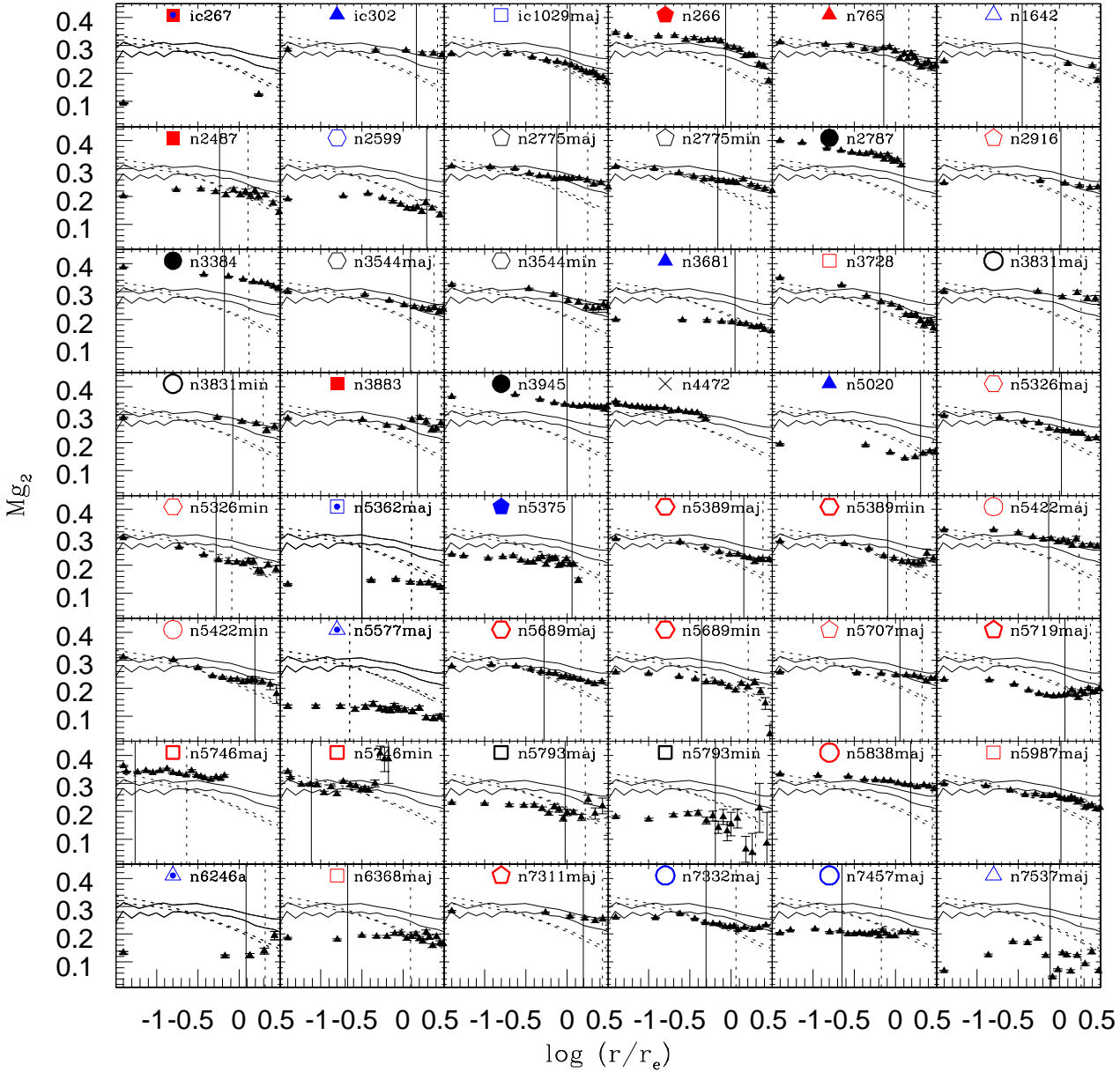}
\caption{Comparison of Mg$_2$ profiles between this work and numerical simulations.  
Solid lines are merger models by Bekki \& Shioya (1999) final age 13.1 Gyr 
and initial disk masses of $10^{10} M_\odot$ (bottom curve) and $10^{12} M_\odot$.  
Dotted lines are collapse models by Angeletti \& Giannone with final age 13 Gyr (top 
curve) and 2 Gyr.  Symbols are as in Fig. 
\ref{hbmetctr}.}\label{mg2comp}\end{figure*}

\begin{figure*}
\includegraphics[width=\textwidth]{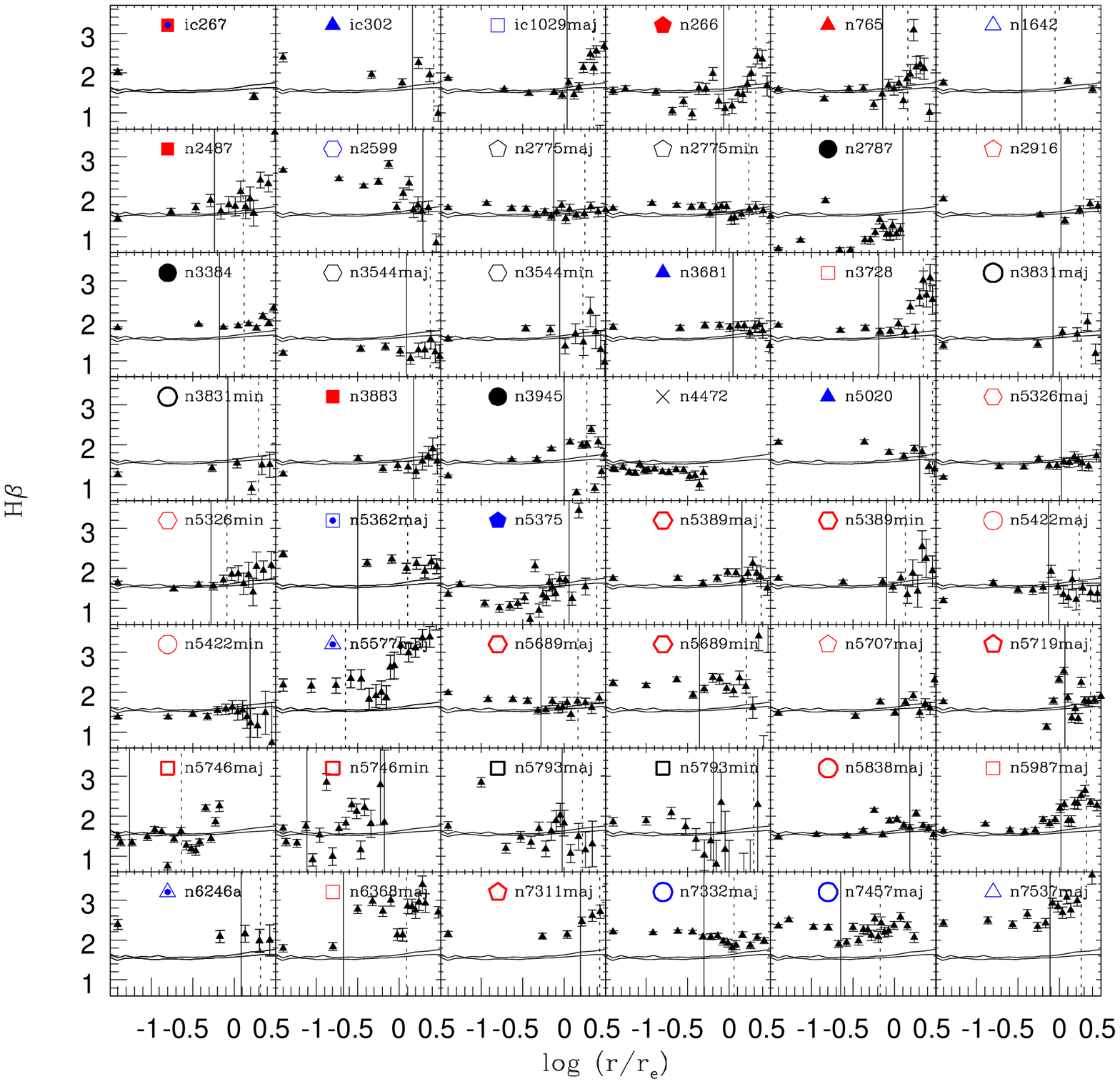}
\caption{Comparison of H$\beta$ profiles between this work and merger models by 
Bekki \& Shioya (1999) with initial disk masses of $10^{10} M_\odot$ (bottom 
curve) and $10^{12} M_\odot$.  Symbols are as in Fig. 
\ref{hbmetctr}.}\label{hbcomp}\end{figure*}

However, the collapse model continues to receive much attention under the claim that 
it better reproduces the observed line strength profiles of ellipticals.  We 
investigate whether or not this is true for bulges.  Gradients in the index Mg$_2$ 
have been computed in galaxies formed in collapse and merger simulations, allowing 
for direct comparisons with our data (Figs. \ref{mg2comp}).  Points 
are our data.  Solid lines are two remnants from major disk-disk mergers by 
\cite{1999ApJ...513..108B} with initial disk masses of $10^{10} M_\odot$ (bottom 
curve in Mg$_2$; top curve in H$\beta$) and $10^{12} M_\odot$.  The remnants are 13.1 
Gyr old.  Dotted lines are two collapse models by \cite{2003A&A...403..449A} with 
final ages of 13 Gyr (top curve) and 2 Gyr.  We focus on the gradient slopes, 
assuming that changes in mass and formation epoch mainly shift the models up or 
down.  The collapse models predict steeper Mg$_2$ profiles than the merger models.  
Within the bulge-dominated region, some of our profiles agree with one of the merger 
models (e.g. NGCs 765 and 7311) while others agree with one of the collapse models 
(e.g. 3728 and the minor axes of NGCs 5326 and 5422).  There are also cases, mostly 
among blue bulges, where the observed profile is flatter than the merger models (e.g. 
IC 302).  However, the majority of galaxies fall between the collapse and merger 
models.

Bekki \& Shioya also computed H$\beta$ profiles in their models (Fig. \ref{hbcomp}). 
Nearly all our galaxies have flat H$\beta$ profiles within the bulge-dominated region 
as predicted by the models.  The profiles of the oldest bulges agree with the 
models in their zero-points as well, while younger bulges lie above the models.

Through chemical evolution modeling, Thomas et al. (1999) studied $\alpha$/Fe ratios 
in ellipticals that formed through a fast ($\tilde 1$ Gyr) collapse of 
star-forming clumps and through mergers of MW-like spirals.  The main difference 
between the two models was that the merging spirals had several Gyr of Fe 
enrichment while the gas involved in the collapse was not pre-enriched in Fe.  
The large central $\alpha$/Fe ratios found in massive ellipticals were reproduced in 
the collapse model.  The merger model does not produce super-solar $\alpha$/Fe 
assuming a Salpeter IMF unless the merger happened early, before the progenitors 
acquired much Fe.  Metallicity and $\alpha$/Fe are anticorrelated in the collapse 
model; since ellipticals have negative gradients in metallicity, the model predicts 
that they should have a positive gradient in $\alpha$/Fe.  The merger model produces 
solar $\alpha$/Fe in the outer regions.  Therefore, uniformly solar $\alpha$/Fe is 
consistent with the merger model while positive gradients are consistent with the 
collapse model.  Pipino et al. (2005) also find positive $\alpha$/Fe 
gradients in a model where ellipticals are formed through the infall of gaseous 
lumps.  Since our blue bulges have uniformly solar $\alpha$/Fe, they are consistent 
with the predictions of the merger model.  Several of the red bulges have positive 
gradients as predicted by the collapse model.  Most of the remaining bulges have 
uniformly super-solar $\alpha$/Fe, which is difficult to reproduce in any of the 
models.

In summary, both collapse and merger models have limited success in reproducing the 
line strength profiles of individual galaxies but neither explains the full range of 
behaviors seen in the data.  It is important to note that hierarchical models are 
only beginning to make robust predictions for SPs, successfully reproducing 
properties traditionally thought to favor the collapse model, such as the 
mass-metallicity relation.   As advancements continue to be made in incorporating gas 
dynamics, star formation, and chemical evolution in cosmologically-motivated merger 
models, it will be interesting to see if the line strength profiles will be 
reproduced as well.

\subsection{Mergers}

In a recent paper, \cite{2005astro.ph..6044F} argued that massive red 
ellipticals could not have formed entirely through major mergers of gas-rich 
components or through dry mergers but through a combination of the two.  Ellipticals 
of the same mass and color could have formed in different ways: through early 
gas-rich mergers of low-mass objects followed by dry mergers or through recent 
gas-rich mergers of more massive objects.  Objects that arrived on the red sequence 
early-on and have been gaining mass through dry mergers will have larger SSP ages 
than those that have arrived on the red sequence near their present mass as the 
result of recent gas-rich mergers.  The former will also have smaller metallicities 
since their last gas-rich mergers were of lower mass progenitors with correspondingly 
lower metallicities according to the gas-phase mass-metallicity relation 
\citep{2003ApJ...599.1006K, 2004ApJ...613..898T}.  The predicted anticorrelation 
between age and metallicity at fixed $\sigma_0$ is seen in ellipticals.  If 
such an anticorrelation exists for bulges, it is not nearly as tight as that of 
ellipticals.  This suggests that an additional formation mechanism might be required 
to explain the SPs of bulges.

Recent semi-analytic models incorporating the Millenium Simulation of cosmic 
structure growth find a correlation between stellar metallicity and stellar mass, 
with the most massive galaxies having roughly solar metallicity 
\citep{2005astro.ph..9725D}.  This is qualitatively consistent with the observed 
[MgFe]'-$\sigma_0$ and [MgFe]'-V$_{max}$ relations.  Massive ellipticals and 
bulges have super-solar central metallicity which is in apparent contradiction with 
de Lucia et al.'s results.  However, they also have negative metallicity gradients.  
The arrows in Fig. \ref{hdamet}, which extend to approximately the bulge effective 
radius, fall around solar metallicity in the massive red bulges.  The metallicity at 
the effective radius is more representative, than the central value, of the mean 
metallicity.  Therefore, the data are not inconsistent with the models.

de Lucia et al. also find in their simulations that less massive ellipticals had more 
extended star formation histories than their massive counterparts.  This is 
consistent with the observed $\alpha$/Fe-$\sigma_0$ and $\alpha$/Fe-V$_{max}$ 
correlations.  These correlations can be produced in starbursts induced by gas-rich 
mergers.  During the starburst, SN II enrich the ISM with $\alpha$-elements.  If 
star-formation is somehow quenched before SN Ia contribute much Fe, the $\alpha$/Fe 
ratio increases.  Dry mergers would add scatter to the $\alpha$/Fe-$\sigma_0$ and 
$\alpha$/Fe-V$_{max}$ relations since they increase $\sigma_0$ and V$_{max}$ without 
altering the $\alpha$/Fe ratio.

The differences in between blue and red bulges at fixed $\sigma_0$ can 
also be explained by mergers.  At fixed $\sigma_0$, blue bulges have smaller SSP ages 
than their red counterparts which suggests that they have undergone gas-rich mergers 
more recently.  The progenitors of the blue bulges have then had more time to acquire 
Fe.  Since the progenitors have small $\alpha$/Fe ratios, so do the remnants.  This 
is seen in the simulations by Thomas et al. (1999), who found that gas-rich mergers 
cannot produce large $\alpha$/Fe ratios unless they happened early in the chemical 
evolution of the progenitors.

\subsection{Secular Evolution}

In dissipationless secular evolution, the bulge is formed through the vertical and 
radial redistribution of disk stars.  In this process, existing gradients can either 
become amplified since the resulting [pseudo]bulge has a smaller scale length than 
the progenitor disk or erased as a consequence of disk heating. But if the disk has 
no gradient, neither should the bulge.  This process cannot be ruled out on the basis 
of [MgFe]' gradients since the majority of galaxies show negative gradients in the 
disk-dominated region.  However, the majority of red bulges have solar $\alpha$/Fe in 
the disk-dominated region despite having super-solar $\alpha$/Fe in the bulge.  
Therefore they could not have been produced through purely dissipationless secular 
evolution.  If secular evolution with gas infall has been responsible for the 
formation of these objects, the star-formation timescales must have been identical 
(at fixed $\sigma_0$) in this scenario as in merger-induced star-formation since red 
bulges and ellipticals follow the same $\alpha$/Fe-$\sigma_0$ relation.  Furthermore, 
the star formation must have been completed several Gyr ago since red bulges have 
large SSP ages.  This goes for the three barred S0s with disk-like structural and 
kinematical properties (NGCs 2787, 3384, and 3945) as well.  These objects are 
identical to ellipticals of comparable $\sigma_0$ in their stellar populations and 
two of them have among the largest central SSP ages observed.

Note that the $\alpha$/Fe ratios of the blue bulges are consistent with 
dissipationless secular evolution.  Unfortunately, neither mergers nor secular 
evolution can be ruled out for blue bulges on the basis of their $\alpha$/Fe ratios.

Secular evolution with gas infall is supported by the frequency of barred 
galaxies with age gradients.  Of the ten galaxies whose central regions are younger 
than the outer regions, seven are barred and one have a b/p bulge.  Bar-driven gas 
infall could lead to extended star-formation in the central region, producing the 
observed age gradient.

If bars are long-lived and the chemical imprints of secular evolution are 
different from those of mergers, we would expect the bulges of barred galaxies to 
have different abundance patterns than those of unbarred galaxies.  We see hints of 
such differences in index-$\sigma_0$ and index-V$_{max}$ relations.  At fixed 
$\sigma_0$ and V$_{max}$, barred galaxies appear to have larger central 
metallicities.

The metallicities of bulges and their disks are correlated.  This is naturally 
explained in processes that involve the bulge being formed from the disk.  However, 
this correlation holds for all galaxies, not just those with bars, blue bulges, or 
bulges identified as having disk-like structure and kinematics.  Therefore either all 
bulges formed secularly and some had their bars destroyed or the other bulge/disk 
formation mechanisms also produce this correlation.

\subsubsection{Evolution of Galaxy Populations}

Small-$\sigma$ bulges fall into two categories: YMR bulges with little or no star 
formation and MP bulges which are actively forming stars.  This suggests that the 
MP bulges would have migrated to the YMR region by the time their star-formation is 
quenched.  Will this be the scenario for all metal-poor bulges (including that of 
the Milky Way) or is the observed anticorrelation between emission strength and 
metallicity the result of small number statistics?  Are there really no metal-poor 
bulges that do not have emission?  Extending this type of study to large galaxy 
samples should shed light into the evolution of small-$\sigma$ bulges.

While all five of the MP galaxies are late-types (Sb-Sc), three of the 
seven YMR galaxies are early types (S0-Sa).  Perhaps, the mechanisms that trigger and 
quench the star-formation are also responsible for tranforming galaxies from late- to 
early-types.  As the YMR bulges age, they will move down to the OMR region.

\section{Summary}

We have studied line strengths in the bulges and inner disks of 38 galaxies in the 
local universe.  Our galaxies span a wide range of Hubble types, central velocity 
dispersions, maximum disk rotational velocities, and inclinations.  The 
low-inclination galaxies include barred and unbarred objects; the edge-on galaxies 
include those with and without boxy/peanut-shaped bulges.  We included several 
galaxies whose bulges were previously identified as being disk-like in their colors 
or kinematics to see if their spectral properties reveal evidence for secular 
evolution.  We use the [MgFe]' index and five Balmer indices to characterize the 
luminosity-weighted metallicities and ages of the SPs and the 
Mgb/$\langle$Fe$\rangle$ index to characterize the $\alpha$/Fe ratios.  Our main 
results are the following:

\begin{itemize}

\item The central regions of bulges range in SSP metallicity from [Z/H]=-0.8 to +0.7 
dex and in SSP age from less than 2 to greater than 15 Gyr.  

\item The central ages and metallicities are sensitive to bulge color which is in 
turn sensitive to central velocity dispersion and maximum disk rotational 
velocity.  

\item Red bulges of all Hubble types are similar to luminous ellipticals in their 
central SPs.  They have large SSP ages and are super-solar in SSP metallicity and 
$\alpha$/Fe.

\item Blue bulges can be separated into two classes: a metal-poor class that is 
restricted to late-types with small velocity dispersion and a young, metal-rich 
class that includes all Hubble types and velocity dispersions.  The metal-poor blue 
bulges are actively forming stars while the metal-rich ones are not.  Low-luminosity 
ellipticals exhibit a similar range of SSP ages and metallicities as blue bulges.

\item Luminous ellipticals and the different types of bulges form a continuous 
and overlapping sequence on diagrams of metallicity- and age-sensitive indices versus 
$\sigma_0$.  At fixed $\sigma_0$, there is no systematic difference between bulges 
and ellipticals on these diagrams but bulges exhibit larger scatter.  At fixed 
$\sigma_0$, age and metallicity are more tightly anticorrelated in ellipticals 
than in bulges.

\item $\alpha$/Fe in red bulges is correlated with $\sigma_0$ and V$_{max}$.  Red 
bulges and ellipticals follow the same $\alpha$/Fe-$\sigma_0$ relation.

\item Most blue bulges (11 out of 14) are consistent with having solar $\alpha$/Fe.  
At fixed $\sigma_0$, blue bulges have smaller $\alpha$/Fe than red bulges and 
ellipticals.

\item Barred galaxies appear to have larger central metallicities than unbarred 
galaxies of the same $\sigma_0$ and V$_{max}$.

\item Most galaxies show a steady decrease in metallicity-sensitive indices with 
radius.  The slope of the gradient is correlated with the central value and 
therefore with the global kinematics.  The bulge- and disk-dominated regions are 
distinct in their line strength profiles, with the disks generally having shallower 
slopes.  The smallest bulges do not have negative line strength gradients; some of 
these have flat profiles in the central region while others have positive gradients.

\item There is a correlation between [MgFe]' strength in the bulge and the disk.  
This correlation holds for all galaxies, not just those with bars, blue bulges, or 
bulges identified as having disk-like structural or kinematical properties.

\item Where positive age gradients (with the central regions being younger) are 
present, they are invariably in barred galaxies.  This suggests that bar-driven star 
formation has occurred.  However, several red bulges in barred galaxies have large 
central SSP ages (although it could be younger than the outer regions) which means 
there has been no significant bar-driven star formation for several Gyr.

\item Four galaxies have super-solar $\alpha$/Fe in the disk-dominated 
region.  The rest are consistent with having solar $\alpha$/Fe in the disk.

\item Objects identified as having disk-like structural or kinematic properties 
do not have noticeably different SPs than other bulges.  They follow the same scaling 
relations as the red bulges and ellipticals and have metallicity gradients.  The 
three barred S0s identified as having bulges with disk-like structural and kinematic 
properties are also $\alpha$-enhanced and therefore do not resemble the majority 
of the disks, including the MW disk at the solar neighborhood.

\item Color profiles agree frequently but not always with line strength profiles.  
Where there is a discrepancy, due likely to the colors being affected by dust, it is 
usually in the central regions.  Consequently, color gradients (computed as the 
difference in color between the center and a characteristic scale length) do not 
necessarily correlate with [MgFe]' gradients, illustrating the value of spectroscopy.

\end{itemize}

Overall, our results are consistent with the hypothesis that mergers have been the 
dominant mechanism responsible for the formation of bulges.  However, some of the 
observations, such as the correlation between bulge and disk metallicity, pose 
significant challenges to the merger scenario.  Furthermore, the possibility that 
barred galaxies follow different scaling relations than unbarred galaxies and are 
overrepresented among galaxies with age gradients supports the secular evolution 
picture.

Central line strengths on a statistically significant sample of ellipticals and 
bulges of barred and unbarred spirals would be invaluable in determining 
whether more than one formation mechanism is required for bulges.  The necessary data 
are already available in the databases of large surveys such as the SDSS.  Spatially 
resolved studies on a smaller, representative sample, would allow for better 
comparisons between gradients in different types of galaxies.

\section{Acknowledgements}

It is a pleasure to thank Anatoly Klypin, Jason Peterson, Jesus Falc{\'o}n-Barroso, 
Reynier Peletier, Claudia Maraston, Daniel Thomas, Scott Trager, and Guy Worthey for 
helpful discussions or assistance with the data analysis.  Partial
funding was provided by the New Mexico Space Grant Consortium.  This research has 
made use of the NASA/IPAC Extragalactic Database (NED) which is operated by the Jet 
Propulsion Laboratory, California Institute of Technology, under contract with the 
National Aeronautics and Space Administration.  

\bibliography{moorthy}

\def\apj{ApJ}
\def\apjl{ApJ}
\def\apjs{ApJS}
\def\apss{AP\&SS}
\def\aj{AJ}
\def\aap{A\&A}
\def\aaps{A\&AS}
\def\pasp{PASP}
\def\mnras{MNRAS}
\def\araa{ARA\&A}

\begin{thebibliography}{}

\bibitem[\protect\citeauthoryear{{Abadi}, {Navarro}, {Steinmetz} \&
  {Eke}}{{Abadi} et~al.}{2003}]{2003ApJ...597...21A}
{Abadi} M.~G.,  {Navarro} J.~F.,  {Steinmetz} M.,    {Eke} V.~R.,  2003, \apj,
  597, 21

\bibitem[\protect\citeauthoryear{{Angeletti} \& {Giannone}}{{Angeletti} \&
  {Giannone}}{2003}]{2003A&A...403..449A}
{Angeletti} L.,  {Giannone} P.,  2003, \aap, 403, 449

\bibitem[\protect\citeauthoryear{{Arimoto} \& {Yoshii}}{{Arimoto} \&
  {Yoshii}}{1987}]{1987A&A...173...23A}
{Arimoto} N.,  {Yoshii} Y.,  1987, \aap, 173, 23

\bibitem[\protect\citeauthoryear{{Aronica}, {Athanassoula}, {Bureau}, {Bosma},
  {Dettmar}, {Vergani} \& {Pohlen}}{{Aronica}
  et~al.}{2003}]{2003Ap&SS.284..753A}
{Aronica} G.,  {Athanassoula} E.,  {Bureau} M.,  {Bosma} A.,  {Dettmar} R.-J.,
  {Vergani} D.,    {Pohlen} M.,  2003, \apss, 284, 753

\bibitem[\protect\citeauthoryear{{Athanassoula}}{{Athanassoula}}{2005}]{2005MN%
RAS.358.1477A}
{Athanassoula} E.,  2005, \mnras, 358, 1477

\bibitem[\protect\citeauthoryear{{Athanassoula} \& {Misiriotis}}{{Athanassoula}
  \& {Misiriotis}}{2002}]{2002MNRAS.330...35A}
{Athanassoula} E.,  {Misiriotis} A.,  2002, \mnras, 330, 35

\bibitem[\protect\citeauthoryear{{Balcells}, {Graham},
  {Dom{\'{\i}}nguez-Palmero} \& {Peletier}}{{Balcells}
  et~al.}{2003}]{2003ApJ...582L..79B}
{Balcells} M.,  {Graham} A.~W.,  {Dom{\'{\i}}nguez-Palmero} L.,    {Peletier}
  R.~F.,  2003, \apjl, 582, L79

\bibitem[\protect\citeauthoryear{{Balcells} \& {Peletier}}{{Balcells} \&
  {Peletier}}{1994}]{1994AJ....107..135B}
{Balcells} M.,  {Peletier} R.~F.,  1994, \aj, 107, 135

\bibitem[\protect\citeauthoryear{{Baldwin}, {Phillips} \&
  {Terlevich}}{{Baldwin} et~al.}{1981}]{1981PASP...93....5B}
{Baldwin} J.~A.,  {Phillips} M.~M.,    {Terlevich} R.,  1981, \pasp, 93, 5

\bibitem[\protect\citeauthoryear{{Bekki} \& {Shioya}}{{Bekki} \&
  {Shioya}}{1999}]{1999ApJ...513..108B}
{Bekki} K.,  {Shioya} Y.,  1999, \apj, 513, 108

\bibitem[\protect\citeauthoryear{{Bell}, {Wolf}, {Meisenheimer}, {Rix},
  {Borch}, {Dye}, {Kleinheinrich}, {Wisotzki} \& {McIntosh}}{{Bell}
  et~al.}{2004}]{2004ApJ...608..752B}
{Bell} E.~F.,  {Wolf} C.,  {Meisenheimer} K.,  {Rix} H.-W.,  {Borch} A.,  {Dye}
  S.,  {Kleinheinrich} M.,  {Wisotzki} L.,    {McIntosh} D.~H.,  2004, \apj,
  608, 752

\bibitem[\protect\citeauthoryear{{Bender}, {Burstein} \& {Faber}}{{Bender}
  et~al.}{1993}]{1993ApJ...411..153B}
{Bender} R.,  {Burstein} D.,    {Faber} S.~M.,  1993, \apj, 411, 153

\bibitem[\protect\citeauthoryear{{Bertola} \& {Capaccioli}}{{Bertola} \&
  {Capaccioli}}{1977}]{1977ApJ...211..697B}
{Bertola} F.,  {Capaccioli} M.,  1977, \apj, 211, 697

\bibitem[\protect\citeauthoryear{{Bournaud}, {Combes} \& {Semelin}}{{Bournaud}
  et~al.}{2005}]{2005MNRAS.tmpL..89B}
{Bournaud} F.,  {Combes} F.,    {Semelin} B.,  2005, \mnras, pp~L89+

\bibitem[\protect\citeauthoryear{{Bruzual} \& {Charlot}}{{Bruzual} \&
  {Charlot}}{2003}]{2003MNRAS.344.1000B}
{Bruzual} G.,  {Charlot} S.,  2003, \mnras, 344, 1000

\bibitem[\protect\citeauthoryear{{Bureau} \& {Freeman}}{{Bureau} \&
  {Freeman}}{1999}]{1999AJ....118..126B}
{Bureau} M.,  {Freeman} K.~C.,  1999, \aj, 118, 126

\bibitem[\protect\citeauthoryear{{Burstein}, {Faber}, {Gaskell} \&
  {Krumm}}{{Burstein} et~al.}{1984}]{1984ApJ...287..586B}
{Burstein} D.,  {Faber} S.~M.,  {Gaskell} C.~M.,    {Krumm} N.,  1984, \apj,
  287, 586

\bibitem[\protect\citeauthoryear{{Busarello}, {Capaccioli}, {D'Onofrio},
  {Longo}, {Richter} \& {Zaggia}}{{Busarello}
  et~al.}{1996}]{1996A&A...314...32B}
{Busarello} G.,  {Capaccioli} M.,  {D'Onofrio} M.,  {Longo} G.,  {Richter} G.,
    {Zaggia} S.,  1996, \aap, 314, 32

\bibitem[\protect\citeauthoryear{{Caldwell}, {Rose} \& {Concannon}}{{Caldwell}
  et~al.}{2003}]{2003AJ....125.2891C}
{Caldwell} N.,  {Rose} J.~A.,    {Concannon} K.~D.,  2003, \aj, 125, 2891

\bibitem[\protect\citeauthoryear{{Cappellari} \& {Emsellem}}{{Cappellari} \&
  {Emsellem}}{2004}]{2004PASP..116..138C}
{Cappellari} M.,  {Emsellem} E.,  2004, \pasp, 116, 138

\bibitem[\protect\citeauthoryear{{Carlberg}}{{Carlberg}}{1984}]{1984ApJ...286.%
.403C}
{Carlberg} R.~G.,  1984, \apj, 286, 403

\bibitem[\protect\citeauthoryear{{Carollo}, {Danziger} \& {Buson}}{{Carollo}
  et~al.}{1993}]{1993MNRAS.265..553C}
{Carollo} C.~M.,  {Danziger} I.~J.,    {Buson} L.,  1993, \mnras, 265, 553

\bibitem[\protect\citeauthoryear{{Carollo}, {Stiavelli}, {de Zeeuw} \&
  {Mack}}{{Carollo} et~al.}{1997}]{1997AJ....114.2366C}
{Carollo} C.~M.,  {Stiavelli} M.,  {de Zeeuw} P.~T.,    {Mack} J.,  1997, \aj,
  114, 2366

\bibitem[\protect\citeauthoryear{{Chung} \& {Bureau}}{{Chung} \&
  {Bureau}}{2004}]{2004AJ....127.3192C}
{Chung} A.,  {Bureau} M.,  2004, \aj, 127, 3192

\bibitem[\protect\citeauthoryear{{Combes}, {Debbasch}, {Friedli} \&
  {Pfenniger}}{{Combes} et~al.}{1990}]{1990A&A...233...82C}
{Combes} F.,  {Debbasch} F.,  {Friedli} D.,    {Pfenniger} D.,  1990, \aap,
  233, 82

\bibitem[\protect\citeauthoryear{{Courteau}, {de Jong} \& {Broeils}}{{Courteau}
  et~al.}{1996}]{1996ApJ...457L..73C}
{Courteau} S.,  {de Jong} R.~S.,    {Broeils} A.~H.,  1996, \apjl, 457, L73

\bibitem[\protect\citeauthoryear{{Davidge}}{{Davidge}}{2001}]{2001AJ....122.13%
86D}
{Davidge} T.~J.,  2001, \aj, 122, 1386

\bibitem[\protect\citeauthoryear{{de Jong}}{{de
  Jong}}{1996}]{1996A&A...313..377D}
{de Jong} R.~S.,  1996, \aap, 313, 377

\bibitem[\protect\citeauthoryear{{de Jong}, {Simard}, {Davies}, {Saglia},
  {Burstein}, {Colless}, {McMahan} \& {Wegner}}{{de Jong}
  et~al.}{2004}]{2004MNRAS.355.1155D}
{de Jong} R.~S.,  {Simard} L.,  {Davies} R.~L.,  {Saglia} R.~P.,  {Burstein}
  D.,  {Colless} M.,  {McMahan} R.,    {Wegner} G.,  2004, \mnras, 355, 1155

\bibitem[\protect\citeauthoryear{{De Lucia}, {Springel}, {White}, {Croton} \&
  {Kauffmann}}{{De Lucia} et~al.}{2005}]{2005astro.ph..9725D}
{De Lucia} G.,  {Springel} V.,  {White} S.~D.~M.,  {Croton} D.,    {Kauffmann}
  G.,  2005, ArXiv e-prints (astro-ph/0509725)

\bibitem[\protect\citeauthoryear{{de Vaucouleurs}, {de Vaucouleurs}, {Corwin},
  {Buta}, {Paturel} \& {Fouque}}{{de Vaucouleurs}
  et~al.}{1991}]{1991trcb.book.....D}
{de Vaucouleurs} G.,  {de Vaucouleurs} A.,  {Corwin} H.~G.,  {Buta} R.~J.,
  {Paturel} G.,    {Fouque} P.,  1991, {Third Reference Catalogue of Bright
  Galaxies}.
Volume 1-3, XII, 2069 pp.~7 figs..~ Springer-Verlag Berlin Heidelberg New York

\bibitem[\protect\citeauthoryear{{Debattista}, {Carollo}, {Mayer} \&
  {Moore}}{{Debattista} et~al.}{2004}]{2004ApJ...604L..93D}
{Debattista} V.~P.,  {Carollo} C.~M.,  {Mayer} L.,    {Moore} B.,  2004, \apjl,
  604, L93

\bibitem[\protect\citeauthoryear{{Denicol{\' o}}, {Terlevich}, {Terlevich},
  {Forbes} \& {Terlevich}}{{Denicol{\' o}} et~al.}{2005}]{2005MNRAS.358..813D}
{Denicol{\' o}} G.,  {Terlevich} R.,  {Terlevich} E.,  {Forbes} D.~A.,
  {Terlevich} A.,  2005, \mnras, 358, 813

\bibitem[\protect\citeauthoryear{{Denicol{\' o}}, {Terlevich}, {Terlevich},
  {Forbes}, {Terlevich} \& {Carrasco}}{{Denicol{\' o}}
  et~al.}{2005}]{2005MNRAS.356.1440D}
{Denicol{\' o}} G.,  {Terlevich} R.,  {Terlevich} E.,  {Forbes} D.~A.,
  {Terlevich} A.,    {Carrasco} L.,  2005, \mnras, 356, 1440

\bibitem[\protect\citeauthoryear{{Eisenstein}, {Hogg}, {Fukugita}, {Nakamura},
  {Bernardi}, {Finkbeiner}, {Schlegel}, {Brinkmann}, {Connolly}, {Csabai},
  {Gunn}, {Ivezi{\' c}}, {Lamb}, {Loveday}, {Munn}, {Nichol}, {Schneider},
  {Strauss}, {Szalay} \& {York}}{2003}]{2003ApJ...585..694E}
{Eisenstein} D.~J.,  {Hogg} D.~W.,  {Fukugita} M.,  {Nakamura} O.,  {Bernardi}
  M.,  {Finkbeiner} D.~P.,  {Schlegel} D.~J.,  {Brinkmann} J.,  {Connolly}
  A.~J.,  {Csabai} I.,  {Gunn} J.~E.,  {Ivezi{\' c}} {\v Z}.,  {Lamb} D.~Q.,
  {Loveday} J.,  {Munn} J.~A.,  {Nichol} R.~C.,  {Schneider} D.~P.,  {Strauss}
  M.~A.,  {Szalay} A.,    {York} D.~G.,  2003, \apj, 585, 694

\bibitem[\protect\citeauthoryear{{Emsellem}, {Cappellari}, {Peletier},
  {McDermid}, {Bacon}, {Bureau}, {Copin}, {Davies}, {Krajnovi{\' c}},
  {Kuntschner}, {Miller} \& {Tim de Zeeuw}}{{Emsellem}
  et~al.}{2004}]{2004MNRAS.352..721E}
{Emsellem} E.,  {Cappellari} M.,  {Peletier} R.~F.,  {McDermid} R.~M.,  {Bacon}
  R.,  {Bureau} M.,  {Copin} Y.,  {Davies} R.~L.,  {Krajnovi{\' c}} D.,
  {Kuntschner} H.,  {Miller} B.~W.,    {Tim de Zeeuw} P.,  2004, \mnras, 352,
  721

\bibitem[\protect\citeauthoryear{{Erwin}, {Beltr{\' a}n}, {Graham} \&
  {Beckman}}{{Erwin} et~al.}{2003}]{2003ApJ...597..929E}
{Erwin} P.,  {Beltr{\' a}n} J.~C.~V.,  {Graham} A.~W.,    {Beckman} J.~E.,
  2003, \apj, 597, 929

\bibitem[\protect\citeauthoryear{{Faber}, {Friel}, {Burstein} \&
  {Gaskell}}{{Faber} et~al.}{1985}]{1985ApJS...57..711F}
{Faber} S.~M.,  {Friel} E.~D.,  {Burstein} D.,    {Gaskell} C.~M.,  1985,
  \apjs, 57, 711

\bibitem[\protect\citeauthoryear{{Faber} et~al.}{2005}]{2005astro.ph..6044F}
{Faber} S.~M.,  {Willmer} C.~N.~A.,  {Wolf} C.,  {Koo} D.~C.,  {Weiner} B.~J.,
  {Newman} J.~A.,  {Im} M.,  {Coil} A.~L.,  {Conroy} C.,  {Cooper} M.~C.,
  {Davis} M.,  {Finkbeiner} D.~P.,  {Gerke} B.~F.,  {Gebhardt} K.,  {Groth}
  E.~J.,  {Guhathakurta} P.,  {Harker} J.,  {Kaiser} N.,  {Kassin} S.,
  {Kleinheinrich} M.,  {Konidaris} N.~P.,  {Lin} L.,  {Luppino} G.,  {Madgwick}
  D.~S.,  {Noeske} K.~M.~K.~G.,  {Phillips} A.~C.,  {Sarajedini} V.~L.,
  {Simard} L.,  {Szalay} A.~S.,  {Vogt} N.~P.,    {Yan} R.,  2005, ArXiv
  e-prints astro-ph/0506044

\bibitem[\protect\citeauthoryear{{Fairall}, {Willmer}, {Calderon}, {Latham},
  {Nicolaci da Costa}, {Pellegrini}, {Nunes}, {Focardi} \&
  {Vettolani}}{{Fairall} et~al.}{1992}]{1992AJ....103...11F}
{Fairall} A.~P.,  {Willmer} C.~N.~A.,  {Calderon} J.~H.,  {Latham} D.~W.,
  {Nicolaci da Costa} L.,  {Pellegrini} P.~S.,  {Nunes} M.~A.,  {Focardi} P.,
   {Vettolani} G.,  1992, \aj, 103, 11

\bibitem[\protect\citeauthoryear{{Falc{\' o}n-Barroso}, {Peletier} \&
  {Balcells}}{{Falc{\' o}n-Barroso} et~al.}{2002}]{2002MNRAS.335..741F}
{Falc{\' o}n-Barroso} J.,  {Peletier} R.~F.,    {Balcells} M.,  2002, \mnras,
  335, 741

\bibitem[\protect\citeauthoryear{{Falc{\' o}n-Barroso}, {Peletier}, {Emsellem},
  {Kuntschner}, {Fathi}, {Bureau}, {Bacon}, {Cappellari}, {Copin}, {Davies} \&
  {de Zeeuw}}{{Falc{\' o}n-Barroso} et~al.}{2004}]{2004MNRAS.350...35F}
{Falc{\' o}n-Barroso} J.,  {Peletier} R.~F.,  {Emsellem} E.,  {Kuntschner} H.,
  {Fathi} K.,  {Bureau} M.,  {Bacon} R.,  {Cappellari} M.,  {Copin} Y.,
  {Davies} R.~L.,    {de Zeeuw} T.,  2004, \mnras, 350, 35

\bibitem[\protect\citeauthoryear{{Feltzing} \& {Gilmore}}{{Feltzing} \&
  {Gilmore}}{2000}]{2000A&A...355..949F}
{Feltzing} S.,  {Gilmore} G.,  2000, \aap, 355, 949

\bibitem[\protect\citeauthoryear{{Ferreiro} \& {Pastoriza}}{{Ferreiro} \&
  {Pastoriza}}{2004}]{2004A&A...428..837F}
{Ferreiro} D.~L.,  {Pastoriza} M.~G.,  2004, \aap, 428, 837

\bibitem[\protect\citeauthoryear{{Fisher}, {Franx} \& {Illingworth}}{{Fisher}
  et~al.}{1996}]{1996ApJ...459..110F}
{Fisher} D.,  {Franx} M.,    {Illingworth} G.,  1996, \apj, 459, 110+

\bibitem[\protect\citeauthoryear{{Forbes}, {S{\' a}nchez-Bl{\' a}zquez} \&
  {Proctor}}{{Forbes} et~al.}{2005}]{2005MNRAS.tmpL..47F}
{Forbes} D.~A.,  {S{\' a}nchez-Bl{\' a}zquez} P.,    {Proctor} R.,  2005,
  \mnras, pp~L47+

\bibitem[\protect\citeauthoryear{{Friedli} \& {Benz}}{{Friedli} \&
  {Benz}}{1995}]{1995A&A...301..649F}
{Friedli} D.,  {Benz} W.,  1995, \aap, 301, 649

\bibitem[\protect\citeauthoryear{{Gadotti} \& {dos Anjos}}{{Gadotti} \& {dos
  Anjos}}{2001}]{2001AJ....122.1298G}
{Gadotti} D.~A.,  {dos Anjos} S.,  2001, \aj, 122, 1298

\bibitem[\protect\citeauthoryear{{Goudfrooij}, {Gorgas} \&
  {Jablonka}}{{Goudfrooij} et~al.}{1999}]{1999Ap&SS.269..109G}
{Goudfrooij} P.,  {Gorgas} J.,    {Jablonka} P.,  1999, \apss, 269, 109

\bibitem[\protect\citeauthoryear{{Ibata} \& {Gilmore}}{{Ibata} \&
  {Gilmore}}{1995}]{1995MNRAS.275..605I}
{Ibata} R.~A.,  {Gilmore} G.~F.,  1995, \mnras, 275, 605

\bibitem[\protect\citeauthoryear{{Idiart}, {de Freitas Pacheco} \&
  {Costa}}{{Idiart} et~al.}{1996}]{1996AJ....112.2541I}
{Idiart} T.~P.,  {de Freitas Pacheco} J.~A.,    {Costa} R.~D.~D.,  1996, \aj,
  112, 2541

\bibitem[\protect\citeauthoryear{{Immeli}, {Samland}, {Gerhard} \&
  {Westera}}{{Immeli} et~al.}{2004}]{2004A&A...413..547I}
{Immeli} A.,  {Samland} M.,  {Gerhard} O.,    {Westera} P.,  2004, \aap, 413,
  547

\bibitem[\protect\citeauthoryear{{Jablonka}, {Courbin}, {Meylan}, {Sarajedini},
  {Bridges} \& {Magain}}{{Jablonka} et~al.}{2000}]{2000A&A...359..131J}
{Jablonka} P.,  {Courbin} F.,  {Meylan} G.,  {Sarajedini} A.,  {Bridges} T.~J.,
     {Magain} P.,  2000, \aap, 359, 131

\bibitem[\protect\citeauthoryear{{Jablonka}, {Gorgas} \&
  {Goudfrooij}}{{Jablonka} et~al.}{2002}]{2002Ap&SS.281..367J}
{Jablonka} P.,  {Gorgas} J.,    {Goudfrooij} P.,  2002, \apss, 281, 367

\bibitem[\protect\citeauthoryear{{Jablonka}, {Martin} \& {Arimoto}}{{Jablonka}
  et~al.}{1996}]{1996AJ....112.1415J}
{Jablonka} P.,  {Martin} P.,    {Arimoto} N.,  1996, \aj, 112, 1415

\bibitem[\protect\citeauthoryear{{Kannappan}, {Jansen} \& {Barton}}{{Kannappan}
  et~al.}{2004}]{2004AJ....127.1371K}
{Kannappan} S.~J.,  {Jansen} R.~A.,    {Barton} E.~J.,  2004, \aj, 127, 1371

\bibitem[\protect\citeauthoryear{{Karachentsev} \& {Makarov}}{{Karachentsev} \&
  {Makarov}}{1996}]{1996AJ....111..794K}
{Karachentsev} I.~D.,  {Makarov} D.~A.,  1996, \aj, 111, 794

\bibitem[\protect\citeauthoryear{{Kauffmann}, {Heckman}, {Tremonti},
  {Brinchmann}, {Charlot}, {White}, {Ridgway}, {Brinkmann}, {Fukugita}, {Hall},
  {Ivezi{\' c}}, {Richards} \& {Schneider}}{{Kauffmann}
  et~al.}{2003}]{2003MNRAS.346.1055K}
{Kauffmann} G.,  {Heckman} T.~M.,  {Tremonti} C.,  {Brinchmann} J.,  {Charlot}
  S.,  {White} S.~D.~M.,  {Ridgway} S.~E.,  {Brinkmann} J.,  {Fukugita} M.,
  {Hall} P.~B.,  {Ivezi{\' c}} {\v Z}.,  {Richards} G.~T.,    {Schneider}
  D.~P.,  2003, \mnras, 346, 1055

\bibitem[\protect\citeauthoryear{{Kauffmann}, {White} \&
  {Guiderdoni}}{{Kauffmann} et~al.}{1993}]{1993MNRAS.264..201K}
{Kauffmann} G.,  {White} S.~D.~M.,    {Guiderdoni} B.,  1993, \mnras, 264, 201+

\bibitem[\protect\citeauthoryear{{Kobayashi} \& {Arimoto}}{{Kobayashi} \&
  {Arimoto}}{1999}]{1999ApJ...527..573K}
{Kobayashi} C.,  {Arimoto} N.,  1999, \apj, 527, 573

\bibitem[\protect\citeauthoryear{{Kobulnicky}, {Willmer}, {Phillips}, {Koo},
  {Faber}, {Weiner}, {Sarajedini}, {Simard} \& {Vogt}}{{Kobulnicky}
  et~al.}{2003}]{2003ApJ...599.1006K}
{Kobulnicky} H.~A.,  {Willmer} C.~N.~A.,  {Phillips} A.~C.,  {Koo} D.~C.,
  {Faber} S.~M.,  {Weiner} B.~J.,  {Sarajedini} V.~L.,  {Simard} L.,    {Vogt}
  N.~P.,  2003, \apj, 599, 1006

\bibitem[\protect\citeauthoryear{{Kormendy}}{{Kormendy}}{1993}]{1993IAUS..153.%
.209K}
{Kormendy} J.,  1993, in IAU Symp. 153: Galactic Bulges {Kinematics of
  extragalactic bulges: evidence that some bulges are really disks}.
pp 209--+

\bibitem[\protect\citeauthoryear{{Kormendy} \& {Illingworth}}{{Kormendy} \&
  {Illingworth}}{1982}]{1982ApJ...256..460K}
{Kormendy} J.,  {Illingworth} G.,  1982, \apj, 256, 460

\bibitem[\protect\citeauthoryear{{Kormendy} \& {Kennicutt}}{{Kormendy} \&
  {Kennicutt}}{2004}]{2004ARA&A..42..603K}
{Kormendy} J.,  {Kennicutt} R.~C.,  2004, \araa, 42, 603

\bibitem[\protect\citeauthoryear{{L{\" u}tticke}, {Dettmar} \& {Pohlen}}{{L{\"
  u}tticke} et~al.}{2000}]{2000A&AS..145..405L}
{L{\" u}tticke} R.,  {Dettmar} R.-J.,    {Pohlen} M.,  2000, \aaps, 145, 405

\bibitem[\protect\citeauthoryear{{Larson}}{{Larson}}{1974}]{1974MNRAS.166..585%
L}
{Larson} R.~B.,  1974, \mnras, 166, 585

\bibitem[\protect\citeauthoryear{{Lee} \& {Worthey}}{{Lee} \&
  {Worthey}}{2005}]{2005ApJS..160..176L}
{Lee} H.-c.,  {Worthey} G.,  2005, \apjs, 160, 176

\bibitem[\protect\citeauthoryear{{MacArthur}}{{MacArthur}}{2005}]{2005ApJ...62%
3..795M}
{MacArthur} L.~A.,  2005, \apj, 623, 795

\bibitem[\protect\citeauthoryear{{MacArthur}, {Courteau} \&
  {Holtzman}}{{MacArthur} et~al.}{2003}]{2003ApJ...582..689M}
{MacArthur} L.~A.,  {Courteau} S.,    {Holtzman} J.~A.,  2003, \apj, 582, 689

\bibitem[\protect\citeauthoryear{{Massey}, {Strobel}, {Barnes} \&
  {Anderson}}{{Massey} et~al.}{1988}]{1988ApJ...328..315M}
{Massey} P.,  {Strobel} K.,  {Barnes} J.~V.,    {Anderson} E.,  1988, \apj,
  328, 315

\bibitem[\protect\citeauthoryear{{Michard} \& {Marchal}}{{Michard} \&
  {Marchal}}{1994}]{1994A&AS..107..187M}
{Michard} R.,  {Marchal} J.,  1994, \aaps, 107, 187

\bibitem[\protect\citeauthoryear{{Minniti}}{{Minniti}}{1996}]{1996ApJ...459..1%
75M}
{Minniti} D.,  1996, \apj, 459, 175

\bibitem[\protect\citeauthoryear{{Nilson}}{{Nilson}}{1973}]{1973ugcg.book.....%
N}
{Nilson} P.,  1973, {Uppsala general catalogue of galaxies}.
Acta Universitatis Upsaliensis.~Nova Acta Regiae Societatis Scientiarum
  Upsaliensis - Uppsala Astronomiska Observatoriums Annaler, Uppsala:
  Astronomiska Observatorium, 1973

\bibitem[\protect\citeauthoryear{{Noguchi}}{{Noguchi}}{2000}]{2000MNRAS.312..1%
94N}
{Noguchi} M.,  2000, \mnras, 312, 194

\bibitem[\protect\citeauthoryear{{Norman}, {Sellwood} \& {Hasan}}{{Norman}
  et~al.}{1996}]{1996ApJ...462..114N}
{Norman} C.~A.,  {Sellwood} J.~A.,    {Hasan} H.,  1996, \apj, 462, 114

\bibitem[\protect\citeauthoryear{{Peletier} \& {Balcells}}{{Peletier} \&
  {Balcells}}{1997}]{1997NewA....1..349P}
{Peletier} R.~F.,  {Balcells} M.,  1997, New Astronomy, 1, 349

\bibitem[\protect\citeauthoryear{{Peletier}, {Vazdekis}, {Arribas}, {del
  Burgo}, {Garc{\' i}a-Lorenzo}, {Guti{\' e}rrez}, {Mediavilla} \&
  {Prada}}{{Peletier} et~al.}{1999}]{1999MNRAS.310..863P}
{Peletier} R.~F.,  {Vazdekis} A.,  {Arribas} S.,  {del Burgo} C.,  {Garc{\'
  i}a-Lorenzo} B.,  {Guti{\' e}rrez} C.,  {Mediavilla} E.,    {Prada} F.,
  1999, \mnras, 310, 863

\bibitem[\protect\citeauthoryear{{Pfenniger}}{{Pfenniger}}{1993}]{1993gabu.sym%
p..387P}
{Pfenniger} D.,  1993, in Galactic Bulges, IAU Symposium 153, (H. Dejonghe \&
  H.J. Habing eds.), Kluwer, Dordrecht, p. 387-390 {Delayed Formation of Bulges
  by Dynamical Processes}.
pp 387--390

\bibitem[\protect\citeauthoryear{{Pinkney}, {Gebhardt}, {Bender}, {Bower},
  {Dressler}, {Faber}, {Filippenko}, {Green}, {Ho}, {Kormendy}, {Lauer},
  {Magorrian}, {Richstone} \& {Tremaine}}{{Pinkney}
  et~al.}{2003}]{2003ApJ...596..903P}
{Pinkney} J.,  {Gebhardt} K.,  {Bender} R.,  {Bower} G.,  {Dressler} A.,
  {Faber} S.~M.,  {Filippenko} A.~V.,  {Green} R.,  {Ho} L.~C.,  {Kormendy} J.,
   {Lauer} T.~R.,  {Magorrian} J.,  {Richstone} D.,    {Tremaine} S.,  2003,
  \apj, 596, 903

\bibitem[\protect\citeauthoryear{{Proctor} \& {Sansom}}{{Proctor} \&
  {Sansom}}{2002}]{2002MNRAS.333..517P}
{Proctor} R.~N.,  {Sansom} A.~E.,  2002, \mnras, 333, 517

\bibitem[\protect\citeauthoryear{{Proctor}, {Sansom} \& {Reid}}{{Proctor}
  et~al.}{2000}]{2000MNRAS.311...37P}
{Proctor} R.~N.,  {Sansom} A.~E.,    {Reid} I.~N.,  2000, \mnras, 311, 37

\bibitem[\protect\citeauthoryear{{Prugniel}, {Maubon} \& {Simien}}{{Prugniel}
  et~al.}{2001}]{2001A&A...366...68P}
{Prugniel} P.,  {Maubon} G.,    {Simien} F.,  2001, \aap, 366, 68

\bibitem[\protect\citeauthoryear{{Puzia}, {Perrett} \& {Bridges}}{{Puzia}
  et~al.}{2005}]{2005A&A...434..909P}
{Puzia} T.~H.,  {Perrett} K.~M.,    {Bridges} T.~J.,  2005, \aap, 434, 909

\bibitem[\protect\citeauthoryear{{Puzia}, {Saglia}, {Kissler-Patig},
  {Maraston}, {Greggio}, {Renzini} \& {Ortolani}}{{Puzia}
  et~al.}{2002}]{2002A&A...395...45P}
{Puzia} T.~H.,  {Saglia} R.~P.,  {Kissler-Patig} M.,  {Maraston} C.,  {Greggio}
  L.,  {Renzini} A.,    {Ortolani} S.,  2002, \aap, 395, 45

\bibitem[\protect\citeauthoryear{{Raimann}, {Storchi-Bergmann}, {Bica} \&
  {Alloin}}{{Raimann} et~al.}{2001}]{2001MNRAS.324.1087R}
{Raimann} D.,  {Storchi-Bergmann} T.,  {Bica} E.,    {Alloin} D.,  2001,
  \mnras, 324, 1087

\bibitem[\protect\citeauthoryear{{Raimann}, {Storchi-Bergmann}, {Gonz{\' a}lez
  Delgado}, {Cid Fernandes}, {Heckman}, {Leitherer} \& {Schmitt}}{{Raimann}
  et~al.}{2003}]{2003MNRAS.339..772R}
{Raimann} D.,  {Storchi-Bergmann} T.,  {Gonz{\' a}lez Delgado} R.~M.,  {Cid
  Fernandes} R.,  {Heckman} T.,  {Leitherer} C.,    {Schmitt} H.,  2003,
  \mnras, 339, 772

\bibitem[\protect\citeauthoryear{{Rich}}{{Rich}}{1999}]{1999ASPC..192..215R}
{Rich} R.~M.,  1999, in ASP Conf. Ser. 192: Spectrophotometric Dating of Stars
  and Galaxies {Age and Metallicity of the Bulges of the Milky Way and M31}.
pp 215--+

\bibitem[\protect\citeauthoryear{{Rose}, {Bower}, {Caldwell}, {Ellis},
  {Sharples} \& {Teague}}{{Rose} et~al.}{1994}]{1994AJ....108.2054R}
{Rose} J.~A.,  {Bower} R.~G.,  {Caldwell} N.,  {Ellis} R.~S.,  {Sharples}
  R.~M.,    {Teague} P.,  1994, \aj, 108, 2054

\bibitem[\protect\citeauthoryear{{Ryder}, {Fenner} \& {Gibson}}{{Ryder}
  et~al.}{2005}]{2005MNRAS.358.1337R}
{Ryder} S.~D.,  {Fenner} Y.,    {Gibson} B.~K.,  2005, \mnras, 358, 1337

\bibitem[\protect\citeauthoryear{{Sadler}, {Rich} \& {Terndrup}}{{Sadler}
  et~al.}{1996}]{1996AJ....112..171S}
{Sadler} E.~M.,  {Rich} R.~M.,    {Terndrup} D.~M.,  1996, \aj, 112, 171

\bibitem[\protect\citeauthoryear{{Sarajedini} \& {Jablonka}}{{Sarajedini} \&
  {Jablonka}}{2005}]{2005AJ....130.1627S}
{Sarajedini} A.,  {Jablonka} P.,  2005, \aj, 130, 1627

\bibitem[\protect\citeauthoryear{{Sellwood}}{{Sellwood}}{1993}]{1993IAUS..153.%
.391S}
{Sellwood} J.~A.,  1993, in IAU Symp. 153: Galactic Bulges {Peanut shaped
  bars}.
pp 391--+

\bibitem[\protect\citeauthoryear{{Shen} \& {Sellwood}}{{Shen} \&
  {Sellwood}}{2004}]{2004ApJ...604..614S}
{Shen} J.,  {Sellwood} J.~A.,  2004, \apj, 604, 614

\bibitem[\protect\citeauthoryear{{Sil'chenko}, {Afanasiev}, {Chavushyan} \&
  {Valdes}}{{Sil'chenko} et~al.}{2002}]{2002ApJ...577..668S}
{Sil'chenko} O.~K.,  {Afanasiev} V.~L.,  {Chavushyan} V.~H.,    {Valdes} J.~R.,
   2002, \apj, 577, 668

\bibitem[\protect\citeauthoryear{{Sil'chenko}, {Moiseev}, {Afanasiev},
  {Chavushyan} \& {Valdes}}{{Sil'chenko} et~al.}{2003}]{2003ApJ...591..185S}
{Sil'chenko} O.~K.,  {Moiseev} A.~V.,  {Afanasiev} V.~L.,  {Chavushyan} V.~H.,
    {Valdes} J.~R.,  2003, \apj, 591, 185

\bibitem[\protect\citeauthoryear{{Stephens}, {Frogel}, {DePoy}, {Freedman},
  {Gallart}, {Jablonka}, {Renzini}, {Rich} \& {Davies}}{{Stephens}
  et~al.}{2003}]{2003AJ....125.2473S}
{Stephens} A.~W.,  {Frogel} J.~A.,  {DePoy} D.~L.,  {Freedman} W.,  {Gallart}
  C.,  {Jablonka} P.,  {Renzini} A.,  {Rich} R.~M.,    {Davies} R.,  2003, \aj,
  125, 2473

\bibitem[\protect\citeauthoryear{{Tanvir}, {Ferguson} \& {Shanks}}{{Tanvir}
  et~al.}{1999}]{1999MNRAS.310..175T}
{Tanvir} N.~R.,  {Ferguson} H.~C.,    {Shanks} T.,  1999, \mnras, 310, 175

\bibitem[\protect\citeauthoryear{{Thomas}, {Maraston} \& {Bender}}{{Thomas}
  et~al.}{2003}]{2003MNRAS.339..897T}
{Thomas} D.,  {Maraston} C.,    {Bender} R.,  2003, \mnras, 339, 897

\bibitem[\protect\citeauthoryear{{Thomas}, {Maraston}, {Bender} \& {de
  Oliveira}}{{Thomas} et~al.}{2005}]{2005ApJ...621..673T}
{Thomas} D.,  {Maraston} C.,  {Bender} R.,    {de Oliveira} C.~M.,  2005, \apj,
  621, 673

\bibitem[\protect\citeauthoryear{{Thomas}, {Maraston} \& {Korn}}{{Thomas}
  et~al.}{2004}]{2004MNRAS.351L..19T}
{Thomas} D.,  {Maraston} C.,    {Korn} A.,  2004, \mnras, 351, L19

\bibitem[\protect\citeauthoryear{{Trager}, {Faber}, {Worthey} \& {Gonz{\'
  a}lez}}{{Trager} et~al.}{2000a}]{2000AJ....119.1645T}
{Trager} S.~C.,  {Faber} S.~M.,  {Worthey} G.,    {Gonz{\' a}lez} J.~J.~.,
  2000a, \aj, 119, 1645

\bibitem[\protect\citeauthoryear{{Trager}, {Faber}, {Worthey} \& {Gonz{\'
  a}lez}}{{Trager} et~al.}{2000b}]{2000AJ....120..165T}
{Trager} S.~C.,  {Faber} S.~M.,  {Worthey} G.,    {Gonz{\' a}lez} J.~J.,
  2000b, \aj, 120, 165

\bibitem[\protect\citeauthoryear{{Trager}, {Worthey}, {Faber}, {Burstein} \&
  {Gonzalez}}{{Trager} et~al.}{1998}]{1998ApJS..116....1T}
{Trager} S.~C.,  {Worthey} G.,  {Faber} S.~M.,  {Burstein} D.,    {Gonzalez}
  J.~J.,  1998, \apjs, 116, 1

\bibitem[\protect\citeauthoryear{{Tremonti}, {Heckman}, {Kauffmann},
  {Brinchmann}, {Charlot}, {White}, {Seibert}, {Peng}, {Schlegel}, {Uomoto},
  {Fukugita} \& {Brinkmann}}{{Tremonti} et~al.}{2004}]{2004ApJ...613..898T}
{Tremonti} C.~A.,  {Heckman} T.~M.,  {Kauffmann} G.,  {Brinchmann} J.,
  {Charlot} S.,  {White} S.~D.~M.,  {Seibert} M.,  {Peng} E.~W.,  {Schlegel}
  D.~J.,  {Uomoto} A.,  {Fukugita} M.,    {Brinkmann} J.,  2004, \apj, 613, 898

\bibitem[\protect\citeauthoryear{{Tripicco} \& {Bell}}{{Tripicco} \&
  {Bell}}{1995}]{1995AJ....110.3035T}
{Tripicco} M.~J.,  {Bell} R.~A.,  1995, \aj, 110, 3035

\bibitem[\protect\citeauthoryear{{van Dokkum}}{{van
  Dokkum}}{2005}]{2005astro.ph..6661V}
{van Dokkum} P.~G.,  2005, ArXiv e-prints (astro-ph/0506661)

\bibitem[\protect\citeauthoryear{{van Dokkum}, {Franx}, {Fabricant}, {Kelson}
  \& {Illingworth}}{{van Dokkum} et~al.}{1999}]{1999ApJ...520L..95V}
{van Dokkum} P.~G.,  {Franx} M.,  {Fabricant} D.,  {Kelson} D.~D.,
  {Illingworth} G.~D.,  1999, \apjl, 520, L95

\bibitem[\protect\citeauthoryear{{van Loon}, {Gilmore}, {Omont}, {Blommaert},
  {Glass}, {Messineo}, {Schuller}, {Schultheis}, {Yamamura} \& {Zhao}}{{van
  Loon} et~al.}{2003}]{2003MNRAS.338..857V}
{van Loon} J.~T.,  {Gilmore} G.~F.,  {Omont} A.,  {Blommaert} J.~A.~D.~L.,
  {Glass} I.~S.,  {Messineo} M.,  {Schuller} F.,  {Schultheis} M.,  {Yamamura}
  I.,    {Zhao} H.~S.,  2003, \mnras, 338, 857

\bibitem[\protect\citeauthoryear{{Vazdekis}}{{Vazdekis}}{1999}]{1999ApJ...513.%
.224V}
{Vazdekis} A.,  1999, \apj, 513, 224

\bibitem[\protect\citeauthoryear{{Vazdekis}, {Kuntschner}, {Davies}, {Arimoto},
  {Nakamura} \& {Peletier}}{{Vazdekis} et~al.}{2001}]{2001ApJ...551L.127V}
{Vazdekis} A.,  {Kuntschner} H.,  {Davies} R.~L.,  {Arimoto} N.,  {Nakamura}
  O.,    {Peletier} R.,  2001, \apjl, 551, L127

\bibitem[\protect\citeauthoryear{{Vazdekis}, {Peletier}, {Beckman} \&
  {Casuso}}{{Vazdekis} et~al.}{1997}]{1997ApJS..111..203V}
{Vazdekis} A.,  {Peletier} R.~F.,  {Beckman} J.~E.,    {Casuso} E.,  1997,
  \apjs, 111, 203

\bibitem[\protect\citeauthoryear{{Wegner}, {da Costa}, {Alonso}, {Bernardi},
  {Wilmer}, {Pellegrini}, {Rit{\' e}} \& {Maia}}{{Wegner}
  et~al.}{2000}]{2000cofl.work...62W}
{Wegner} G.,  {da Costa} L.~N.,  {Alonso} M.~V.,  {Bernardi} M.,  {Wilmer}
  C.~N.~A.,  {Pellegrini} P.~S.,  {Rit{\' e}} C.,    {Maia} M.,  2000, in ASP
  Conf. Ser. 201: Cosmic Flows Workshop {The Nearby Early-type Galaxies Survey
  (ENEAR): Project Description and Some Preliminary Results}.
pp 62--+

\bibitem[\protect\citeauthoryear{{Worthey}}{{Worthey}}{1994}]{1994ApJS...95..1%
07W}
{Worthey} G.,  1994, \apjs, 95, 107

\bibitem[\protect\citeauthoryear{{Worthey} \& {Collobert}}{{Worthey} \&
  {Collobert}}{2003}]{2003ApJ...586...17W}
{Worthey} G.,  {Collobert} M.,  2003, \apj, 586, 17

\bibitem[\protect\citeauthoryear{{Worthey}, {Faber}, {Gonzalez} \&
  {Burstein}}{{Worthey} et~al.}{1994}]{1994ApJS...94..687W}
{Worthey} G.,  {Faber} S.~M.,  {Gonzalez} J.~J.,    {Burstein} D.,  1994,
  \apjs, 94, 687

\bibitem[\protect\citeauthoryear{{Worthey} \& {Ottaviani}}{{Worthey} \&
  {Ottaviani}}{1997}]{1997ApJS..111..377W}
{Worthey} G.,  {Ottaviani} D.~L.,  1997, \apjs, 111, 377

\bibitem[\protect\citeauthoryear{{Wyse}, {Gilmore} \& {Franx}}{{Wyse}
  et~al.}{1997}]{1997ARA&A..35..637W}
{Wyse} R.~F.~G.,  {Gilmore} G.,    {Franx} M.,  1997, \araa, 35, 637

\bibitem[\protect\citeauthoryear{{Zoccali}, {Renzini}, {Ortolani}, {Greggio},
  {Saviane}, {Cassisi}, {Rejkuba}, {Barbuy}, {Rich} \& {Bica}}{{Zoccali}
  et~al.}{2003}]{2003A&A...399..931Z}
{Zoccali} M.,  {Renzini} A.,  {Ortolani} S.,  {Greggio} L.,  {Saviane} I.,
  {Cassisi} S.,  {Rejkuba} M.,  {Barbuy} B.,  {Rich} R.~M.,    {Bica} E.,
  2003, \aap, 399, 931

\end{thebibliography}

\clearpage

\begin{table*}
\caption{The Galaxy Sample.  DJ denotes de Jong (1996), PB denotes 
Peletier \& Balcells (1996), and EN denotes the ENEAR survey 
(Wegner et al. 2000). Morphological types are from the NASA/IPAC Extragalactic 
Database.  B magnitudes are from RC3 (de Vaucouleurs et al. 1991).  Bulge and disk 
colors are from DJ and PB.  Bulge color is defined to be the color at half the K-band 
bulge effective radius or 5 arcsec, whichever is larger.  Disk color is defined to be 
the color at two disk scale lengths.  $b/a$ is the red major over minor axis ratio 
taken from the sources listed.  The recessional velocities shown are the RC3 
heliocentric velocities corrected to the Local Group according to 
Karachentsev \& Makarov (1996).  Where both 21cm and optical velocities were 
available, the 21 cm values were used.  Since no RC3 data was available for NGC 3831, 
the B magnitude and optical heliocentric velocity were taken from 
Fairall et al. (1992).  The scale was obtained assuming H$_0$=70.  The distance 
to NGC 3384 was determined by Cepheid observations 
(Tanvir et al. 1999).}\begin{center}\begin{tabular}{lccccccccc}\hline
\multicolumn{1}{l} {Galaxy} &  {Source} & {Type} 
& {Morph.} & {m$_B$ (mag)} & {(B-K)$_B$} & {(B-R)$_D$} & {b/a} & {V$_{LG}$ (km/s)} & 
{Scale (kpc/arcsec)} \\ \hline
IC 267&DJ&SBb&Bar&13.63&4.6&4.24&0.71&3577&0.25\\
IC 302&DJ&SBbc&Bar&13.59&&&0.92&5950&0.41\\
IC 1029&PB&SAb&Ell$^*$&13.64&3.89&3.77&0.24&2520&0.17\\
NGC 266&DJ&SBab&Bar&12.33&4.6&3.85&0.94&4908&0.34\\
NGC 765&DJ&SABbc&Bar&13.60&4.4&3.82&1.00&5117&0.37\\
NGC 1642&DJ&SA(rs)c&Unb?&13.28&4.0&3.26&1.00&4579&0.32\\
NGC 2487&DJ&SBb&Bar&13.10&4.1&3.53&0.92&4758&0.33\\
NGC 2599&DJ&SAa&Unb&13.12&4.0&3.8&1.00&4651&0.32\\
NGC 2775&EN&SAab&Unb&11.13&&&0.85&1173&0.08\\
NGC 2787&&SB0+&Bar&11.77&&&&696&0.06\\
NGC 2916&DJ&SAab&Unb&12.42&4.2&3.59&0.74&3618&0.25\\
NGC 3384&&SB0-&Bar&10.63&&&&735&0.06\\
NGC 3544&EN&SABa&Unb?&12.99&&&0.30&3354&0.23\\
NGC 3681&DJ&SAB(r)bc&Bar&12.25&3.8&3.47&1.00&1239&0.08\\
NGC 3728&DJ&SAb&Unb&13.80&4.2&3.7&0.75&6904&0.48\\
NGC 3831&EN&SAB0+&Box$^*$&14.5&&&0.24&4715&0.33\\
NGC 3883&DJ&SAb&Bar&13.10&4.1&3.44&0.91&6937&0.48\\
NGC 3945&&SB0+&Bar&11.38&&&&1220&0.09\\
NGC 4472&EN&E2&&9.30&&&&744&0.05\\
NGC 5020&DJ&SABbc&Bar&12.50&3.6&3.11&0.85&3284&0.23\\
NGC 5326&PB&SAa&Unb&12.92&4.05&3.97&0.50&2573&0.18\\
NGC 5362&PB&SAb&Unb&13.14&3.56&3.24&0.37&2314&0.16\\
NGC 5375&DJ&SBab&Bar&12.40&3.9&3.47&0.81&2418&0.17\\
NGC 5389&PB&SABO/a&Box$^*$&13.10&4.12&4.10&0.20&1996&0.14\\
NGC 5422&PB&SA0&Ell$^*$&12.81&4.17&4.09&0.20&1921&0.13\\
NGC 5577&PB&SAbc&Ell$^*$&13.05&3.84&3.54&0.28&1702&0.12\\
NGC 5689&PB&SBO/a&Box?$^*$&12.54&4.14&4.12&0.25&2295&0.16\\
NGC 5707&PB&SAab&Ell$^*$&13.38&4.24&3.92&0.25&2354&0.16\\
NGC 5719&PB&SABab&Box?$^*$&13.1&4.54&3.84&0.36&1676&0.12\\
NGC 5746&PB&SABb&Pea$^*$&11.38&4.42&4.50&0.16&1676&0.12\\
NGC 5793&EN&SABb&Box?&14.30&&&0.37&3387&0.23\\
NGC 5838&PB&SA0-&Box?$^*$&11.74&4.21&4.11&0.35&1338&0.09\\
NGC 5987&PB&SAb&Ell&13.00&4.46&4.14&0.40&3207&0.22\\
NGC 6246A&DJ&SABc&Unb&14.10&3.9&3.23&0.91&5495&0.38\\
NGC 6368&PB&SAb&Ell$^*$&13.10&4.84&4.58&0.20&2904&0.20\\
NGC 7311&PB&SAab&Ell&13.36&4.35&4.07&0.50&4762&0.33\\
NGC 7332&PB&SAB0&Box$^*$&12.11&3.75&3.58&0.26&1584&0.11\\
NGC 7457&PB&SAB0-&Box?&11.86&3.69&3.50&0.52&1115&0.08\\
NGC 7537&PB&SAbc&Ell$^*$&13.65&3.88&3.62&0.34&2888&0.20\\

\hline

\end{tabular}
\end{center}
\end{table*}

\clearpage
\begin{table}
\caption{Spectrograph specifications during our observing runs}
\begin{center}
\begin{tabular}{lccccc} \hline
\multicolumn{1}{l} {Detector} & {Obs. Dates (M/D/Y)} & {Grating} & 
{Disp. (\AA/pix)} & {Approx. Resol. (\AA)} & {Scale (Arcsec/pix)}\\\hline
DIS I Blue&1/10/00-2/11/02&Med&3.18&5.7&1.086\\
DIS I Red&1/10/00-2/11/02&Med&3.53&8.6&0.605\\
DIS II Blue&4/13/02-10/09/02&Low&3.05&8.6&0.600\\
DIS II Red&4/13/02-04/07/03&Med&3.13&7.8&0.605\\
DIS III Blue&03/06/03-02/15/04&Low&2.42&7.7&0.419\\
DIS III Red&05/29/03-02/15/04&Med&2.31&6.9&0.396\\

\hline
\end{tabular}
\end{center}
\end{table}

\clearpage
\begin{table}
\caption{Spectroscopic Observations}
\begin{center}
\begin{tabular}{lcccc} \hline
\multicolumn{1}{l} {Galaxy} & {Axis} & {PA} & {Date (M/D/Y)} & {Exp. Time (Sec)} \\
\hline

IC 267&Bar&-25&12/22/03&1x2400\\
&&&&1x1230\\
IC 302&Bar&8&10/9/02&2x2400\\
IC 1029&Maj&152&5/30/03&3x2400\\
NGC 266&Bar&0&9/17/02&2x2400\\
NGC 765&&15&12/22/03&3x2400\\
NGC 1642&&0&12/1/03&2x2400\\
NGC 2487&Bar&45&2/11/02&3x2400\\
NGC 2599&&-90&2/11/02&3x2400\\
NGC 2775&Maj&66&1/10/00&3x1200\\
&Min&156&1/10/00&3x1200\\
NGC 2787&Maj&109&2/15/04&2x2400\\
NGC 2916&Min&-80&12/1/03&2x2400\\
&&&&1x900\\
NGC 3384&Maj&50&2/15/04&2x2400\\
NGC 3544&Maj&-84&1/10/00&3x1200\\
&Min&6&4/25/00&3x1200\\
NGC 3681&Bar&-25&2/11/02&3x2400\\
NGC 3728&Maj&20&3/6/03&3x2400\\
NGC 3831&Maj&24&4/25/00&3x1200\\
&Min&114&5/3/00&3x1800\\
NGC 3883&Maj&-14&3/7/03&3x2400\\
&&&&1x1200\\
NGC 3945&Maj&-22&2/15/04&2x2400\\
&&&&1x1800\\
NGC 4472&&67&1/10/00&3x1200\\     
NGC 5020&Bar&38&3/6/03&2x2400\\
NGC 5326&Maj&-44&5/4/00&4x1800\\
&Min&-134&2/11/02&2x2400\\
NGC 5362&Maj&-92&6/16/01&4x2400\\
NGC 5375&Bar&-10&4/7/03&1x2400\\                            
&&&&1x1200\\
NGC 5389&Maj&3&5/2/00&4x1800\\
&Min&-87&5/2/00&3x1800\\
NGC 5422&Maj&-26&5/2/00&2x1200\\
%&&&5/3/00&1x1800\\
&Min&64&5/3/00&2x1800\\
&&&&2x1500\\
NGC 5577&Maj&56&1/10/00&3x1200\\
&&&&2x1500\\
NGC 5689&Maj&-93&6/17/01&3x2400\\
&Min&0&4/13/02&2x2400\\
NGC 5707&Maj&39&5/29/03&2x2400\\
&&&&1x1200\\
NGC 5719&Maj&-90&5/30/03&3x2400\\
NGC 5746&Maj&-9&4/17/02&3x2400\\
&Min&-99&4/17/02&3x2400\\
NGC 5793&Maj&-35&5/4/00&1x1800\\
&Min&55&&2x1500\\
NGC 5838&Maj&42&6/8/02&2x2400\\
NGC 5987&Maj&-109&5/30/03&1x2400\\
&&&&1x2700\\
NGC 6246A&&-90&2/19/01&2x1800\\
&&&2/20/01&2x1800\\
NGC 6368&Maj&47&6/29/03&1x2400\\
&&&&1x2100\\
NGC 7311&Maj&24&7/1/03&1x2400\\
&&15&10/12/01&1x2400\\
NGC 7332&Maj&-24&7/3/00&4x1800\\
NGC 7457&Maj&-38&1x2400\\
&&&&1x1257\\
NGC 7537&Maj&-100&10/12/01&4x2400\\
\hline
\end{tabular}
\end{center}
\end{table}
\end{document}